\documentclass[aps,prd,reprint,nofootinbib,longbibliography]{revtex4-1}

\usepackage{amsmath}
\usepackage{mathtools}
\usepackage{graphicx}
\usepackage[normalem]{ulem}
\usepackage[com pat=1.1.0]{tikz-feynman}
\usepackage{tikz}
\usepackage{aas_macros}
\usetikzlibrary{external}
\usepackage{comment}
\usetikzlibrary{arrows}
\usepackage[colorlinks=true, allcolors=blue]{hyperref}

\newcommand\barparena[1]{\overset{%
   \scriptscriptstyle(-)}{#1}}

\begin{document}

\title{Basic characteristics of neutrino flavor conversions in the post-shock regions of core-collapse supernova}

\author{Hiroki Nagakura}
\email{hiroki.nagakura@nao.ac.jp}
\affiliation{Division of Science, National Astronomical Observatory of Japan, 2-21-1 Osawa, Mitaka, Tokyo 181-8588, Japan}
\author{Masamichi Zaizen}
\affiliation{Faculty of Science and Engineering, Waseda University, Tokyo 169-8555, Japan}

\begin{abstract}
One of the active debates in core-collapse supernova (CCSN) theory is how significantly neutrino flavor conversions induced by neutrino-neutrino self-interactions change the conventional picture of CCSN dynamics. Recent studies have indicated that strong flavor conversions can occur inside neutrino spheres where neutrinos are tightly coupled to matter. These flavor conversions are associated with either collisional instability or fast neutrino-flavor conversion (FFC) or both. The impact of these flavor conversions on CCSN dynamics is, however, still highly uncertain due to the lack of global simulations of quantum kinetic neutrino transport with appropriate microphysical inputs. Given fluid profiles from a recent CCSN model at three different time snapshots in the early post-bounce phase, we perform global quantum kinetic simulations in spherical symmetry with an essential set of microphysics. We find that strong flavor conversions occur in optically thick regions, resulting in a substantial change of neutrino radiation field. The neutrino heating in the gain region is smaller than the case with no flavor conversions, whereas the neutrino cooling in the optically thick region is commonly enhanced. Based on the neutrino data obtained from our multi-angle neutrino transport simulations, we also assess some representative classical closure relations by applying them to diagonal components of density matrix of neutrinos. We find that Eddington tensors can be well approximated by these closure relations except for the region where flavor conversions occur vividly. We also analyze the neutrino signal by carrying out detector simulations for Super-Kamiokande, DUNE, and JUNO. We propose a useful strategy to identify the sign of flavor conversions in neutrino signal, that can be easily implemented in real data analyses of CCSN neutrinos.
\end{abstract}
\maketitle

\section{Introduction}\label{sec:intro}
One of the lingering mysteries in stellar astrophysics is why massive stars end their lives as catastrophic explosions, known as core-collapse supernova (CCSN). A comprehensive understanding of the final stage of massive stars is crucial for many areas of modern high-energy astrophysics, since the explosion mechanism of CCSN tightly links to their key questions. What fraction of massive stars leaves neutron star (NS) or black hole? What explosion mechanism provides a consistent explanation for the observed distributions of explosion energy, nickel mass (as well as other nucleosynthesis yields), ejecta morphology, pulsar kick/spin, and their mutual correlations? Because of the complex and non-linear nature of CCSN physics, addressing these issues require detailed theoretical models. The next nearby CCSN event will provide precious information to answer these questions, since it may bring us the first simultaneous detection of electromagnetic waves (EM), neutrinos, and gravitational waves (GWs). Theoretical models will be used as a reference to extract physical quantities from these observed signals, indicating that the physical fidelity of CCSN models is of critical importance.

Over the past several decades, significant progress has been made on this front. Nowadays, numerical simulations reproduce, though not all, successful CCSN explosions with sophisticated input physics \cite{2018ApJ...855L...3O,2020MNRAS.498L.109M,2020ApJ...896..102K,2020MNRAS.491.2715B,2021MNRAS.507..443B,2021ApJ...915...28B,2022MNRAS.516.1752M,2023PhRvD.107d3008M,2023MNRAS.522.6070P}. Based on multi-dimensional CCSN models, the sensitivity of CCSN dynamics to the physical fidelity of numerical simulations has also been explored in detail, e.g., for neutrino-matter interactions \cite{2017PhRvL.119x2702B,2018SSRv..214...33B,2021ApJ...906..128K}, equation-of-states (EOS) \cite{2018ApJ...857...13P,2019PhRvC.100e5802S,2021ApJ...923..201E}, self-gravity \cite{2018ApJ...854...63O}, and neutrino transport \cite{2018A&A...619A.118C}. According to these studies, observed signals hinge on the detail of each physical process. This suggests that many unconstrained parameters associated with CCSN physics can be explored with the multi-messenger observation.

Though CCSN models have undergone a significant maturation, crucial open questions still remain to be solved. One of the major uncertainties is neutrino quantum kinetics. In canonical CCSN models, neutrino oscillations are taken into account with vacuum oscillation (driven by neutrino mass differences and mixing angles) and refractive effects from medium. In high-density environments, the refractive effect dominates the neutrino oscillation, which keeps neutrinos in flavor eigenstates. The neutrino kinetic equation becomes, thus, the same form as classical transport one, implying that the time evolution of specific intensity (or equivalently distribution function) in six-dimensional phase space corresponds to the complete information representing the dynamics of neutrino radiation field. For this reason, the neutrino oscillation has been effectively neglected in the study of CCSN explosion mechanism. We note, however, that the neutrino oscillation occurs when neutrinos propagate in the outer envelope of progenitor (since the matter density becomes lower). This effect has usually been taken into account in discussions of neutrino signal observed at the Earth (see, e.g., \cite{2000PhRvD..62c3007D}).

In reality, however, neutrino flavor conversions would be more complicated than what we discussed above. A missing ingredient is neutrino-neutrino self-interactions. The dense neutrino media in CCSN core can generate strong refractive effects by neutrinos themselves. Although the power of refraction is usually several orders of magnitude weaker than the matter one, various types of instability of flavor conversion can happen. One of the noticeable differences from the matter effect is that the refraction by self-interactions acts on not only neutrinos in flavor eigenstates but also those in the coherent states, which leads to unique non-linear evolution of flavor conversions. Though it has been recognized for many years that neutrino oscillations associated with self-interactions can occur in CCSN core, their impacts on CCSNe remain shrouded in mystery, because of the complexities due to the non-linear process, and a huge disparity of physical scales between flavor conversions and astrophysical systems.

During the past few decades, various efforts have been made to understand these flavor instabilities driven by neutrino self-interactions. Normal-mode analysis is a powerful approach to categorize the instabilities \cite{2018JCAP...12..019A}, which is also useful to highlight driving mechanisms of flavor conversions. The so-called slow flavor conversion, which is triggered by the interplay between vacuum oscillation (i.e., neutrino mixing angles and mass differences) and refractive effects by self-interactions, has been investigated in detail in the early 2000's (see a review of \cite{2010ARNPS..60..569D} and references therein). Various phenomena, e.g., neutrino spectral swaps/splits are known to be associated with the slow mode. However, the slow flavor conversion is unlikely to induce significant flavor conversions in the post-shock region \cite{2008PhRvD..78h5012E,2011PhRvL.107o1101C,2011PhRvD..84b5002C,2012PhRvD..85f5008D}, seeming to have little effect on CCSN dynamics (bust see discussions by \cite{2023PhRvD.107f3025S})\footnote{It should be mentioned, however, that the slow mode may be vibrant in the case with small-mass progenitor stars ($\lesssim 9 M_{\odot}$, where $M_{\odot}$ denotes the solar mass) due to lower matter density environments than those in higher mass CCSN progenitors; see, e.g., \cite{2013PhRvD..87h5037C,2020JCAP...06..011Z,2021PhRvD.103f3008Z}.}.

Recent studies have paid attention to two other modes: fast neutrino-flavor conversion (FFC) \cite{2005PhRvD..72d5003S} and collisional instability \cite{2021arXiv210411369J}. They have a potential to induce large flavor conversions even in dense matter environments. The former instability corresponds to one of the collective neutrino oscillation modes, which is driven by anisotropy of neutrino angular distributions in momentum space (see also recent reviews of \cite{2020arXiv201101948T,2022Univ....8...94C,2022arXiv220703561R,2023arXiv230111814V,2023arXiv230803962F}). The growth rate of the instability is proportional to the number density of neutrinos, and the associated time scale of flavor conversion in the vicinity of proto-neutron star (PNS) can be an order of picosecond, that is much shorter than the time scale of neutrino-matter interactions. This indicates that neutrinos may decouple from matter even in optically thick (or diffusion) regions where neutrinos are almost in thermal and chemical equilibrium with matter. This potentially leads to a radical change of neutrino radiation field, and subsequently affects fluid dynamics and nucleosynthesis in CCSNe.

The collisional instability is, on the other hand, driven by neutrino emission, absorption, and scatterings, implying that the collision term is responsible for the instability. The instability may seem counter-intuitive, since neutrino-matter interactions usually cause the decoherence of neutrinos, which suppresses the flavor conversion. However, \citet{2021arXiv210411369J} pointed out that the disparity of decoherence between neutrinos and antineutrinos in neutrino-matter collisions can enhance the flavor coherency, resulting in an acceleration of flavor conversions. The time scale associated with this instability is usually comparable to that of neutrino matter interaction, but it can be significantly shorter in resonance regions \cite{2022arXiv221203750X,2023arXiv230206263L}. This suggests that the collisional instability may change the neutrino radiation field substantially, although the actual impact on CCSN is still a matter of debate; see, e.g., controversial results between \cite{2023PhRvD.107h3016X} and \cite{2023arXiv230710366S}.

Without solving non-linear quantum kinetic neutrino transport equation, we can determine occurrences of FFC and collisional instability from linear stability analysis (see, e.g., \cite{2017PhRvD..96d3016C,2017PhRvL.118b1101I,2018JCAP...12..019A} for FFCs and \cite{2023arXiv230206263L} for collisional instability) or its surrogate approaches \cite{2018PhRvD..98j3001D,2020JCAP...05..027A,2021PhRvD.103l3012J,2021PhRvD.104f3014N,2022PhRvD.106h3005R}. Much efforts have been devoted in the last several years to inspect the flavor conversion instabilities in CCSN core by post-processing analysis of numerical models. Many of these works have suggested that unstable regions are present in the post-shock regions; see, e.g., \cite{2021PhRvD.104h3025N,2021PhRvD.103f3033A} for FFC and \cite{2023PhRvD.107h3016X} for collisional instability.

These positive signs of flavor conversion indicated by stability analyses have motivated the study of flavor conversion in the non-linear phase, and detailed investigations have been carried out so far extensively with local-box simulations; see recent works, e.g., \cite{2020PhRvD.102j3017J,2020PhRvD.102f3018B,2021PhRvL.126f1302B,2021PhRvD.103f3002S,2021PhRvD.103f3001M,2021PhRvD.104j3023R,2021PhRvD.104j3003W,2022PhRvD.105d3005S,2022PhRvD.106j3039B,2023PhRvD.107j3022Z,2023PhRvD.107l3021Z,2023arXiv230711129X} for FFC, \cite{2022PhRvD.106j3029J,2023PhRvD.107h3034L} for collisional instability, and for both of them \cite{2023PhRvD.108b3006K}. Although these local-box simulations are highly valuable to understand the fundamental properties of flavor conversions, they are not sufficient to quantify the impact of flavor conversions on CCSN dynamics. This is simply because neutrino transport is a non-local process, and effects of global advection are usually neglected. We also note that CCSN environments are the overall spherical geometry, implying that geometrical effects play key roles in determining the angular distribution of neutrinos in momentum space. This suggests that one needs to accommodate effects of global advection in local simulations one way or another, otherwise one should carry out global simulations to determine roles of flavor conversion in CCSN theory.

For global simulations, however, some approximate prescriptions need to be adopted, because of many technical issues involved in quantum kinetic neutrino transport. A simple but useful approach is that radiation-hydrodynamic simulations are carried out with classical neutrino transport, while flavor conversions are handled with a parametric manner \cite{2023PhRvD.107j3034E,2023arXiv230511207E}\footnote{Although the detail of prescriptions and parameters are different, the similar approach has been taken in the study of FFC in binary neutron star merger environments \cite{2021PhRvL.126y1101L,2022PhRvD.105h3024J,2022PhRvD.106j3003F}.}. According to these studies, neutrino flavor conversions have strong influence on fluid dynamics and neutrino signal in CCSNe. On the other hand, these studies found that the impact of flavor conversion on CCSN dynamics hinges on the choice of parameters; for instances the degree of flavor conversion and the criterion for the occurrence of flavor instabilities. These parameters can not be determined by these phenomenological models, and solving quantum kinetic equation (QKE) for neutrino transport are the most reliable or perhaps the only way to address this issue.

Motivated by these facts, quantum kinetic neutrino transport simulations have been performed under frozen fluid backgrounds that are set by toy models or taken from numerical CCSN simulations \cite{2022PhRvL.129z1101N,2023PhRvD.107f3025S,2023PhRvD.107f3033N,2023PhRvL.130u1401N,2023PhRvD.107h3016X}. To make these global simulations tractable, we usually reduce the magnitude of neutrino oscillation Hamiltonian, while the impact of the artificial prescription can be studied by convergence studies of the attenuation parameter \cite{2022PhRvL.129z1101N,2023PhRvD.107f3033N}\footnote{We note that \citet{2023PhRvD.107f3025S} did not attenuate the Hamiltonian and directly perform quantum kinetic simulations with $\sim 100$m spatial resolutions.}. All global simulations carried out so far showed rather commonly that FFC and collisional instability substantially change neutrino radiation field in CCSN core. In our recent study \cite{2023PhRvL.130u1401N}, we also show that neutrino heating in the gain region and cooling in the optically thick region are reduced and enhanced by FFCs, respectively. This is an important evidence that fluid dynamics are affected by flavor conversions.

It should be noted that the numerical model in \cite{2023PhRvL.130u1401N} is limited only for a time snapshot of a CCSN model: $300$ ms after core bounce in a CCSN model of \cite{2019ApJS..240...38N}. According to the studies in \cite{2023PhRvD.107j3034E,2023arXiv230511207E}, however, roles of flavor conversions on CCSN may vary with time, in particular at the early post-bounce phase ($< 300$ ms). We need to, hence, expand our previous study to other time snapshots for more comprehensive understanding of roles of FFCs on the neutrino heating mechanism. Another lack of work in our previous study is to analyze the detailed property of flavor conversions. A natural question associated with this issue is to judge whether collisional instability occurs in our model and whether it can overwhelm FFCs. It should be mentioned that we solve QKE with neutrino emission, absorption, and scatterings, indicating that collisional instability may occur in our models. Stability analysis is the most appropriate approach to answer these questions.

Having in mind these unaddressed issues, we present results of quantum kinetic neutrino transport simulations on three fluid profiles at different time snapshots (${\rm T}_{\rm b}=100$, $200$ and $300$ ms, while ${\rm T}_{\rm b}$ denotes the time measured from core bounce) in a numerical CCSN model, and also conduct linear stability analysis to inspect the property of flavor conversion in this study. Based on the obtained neutrino distributions in quasi-steady states, we also quantify Eddington tensors of each species of neutrinos, and compare them to some representative closure relations for diagonal components of density matrix. This analysis provides an important assessment for other approximate methods of quantum kinetic neutrino transport, in particular for two-moment methods (see, e.g., \cite{2022arXiv220702214G}). We also provide neutrino signal with performing detector simulations, and then highlight some qualitative differences from classical neutrino transport.

This paper is organized as follows. In Sec.~\ref{sec:baseq}, we start with a basic equation. Some key information about fluid profiles of our CCSN model, and computational setup of quantum kinetic simulations are provided in Sec.~\ref{sec:model}. We then present our linear stability analysis of flavor conversions for a steady state neutrino radiation field obtained from classical neutrino transport simulations in Sec.~\ref{sec:linana_ini}. All results in our QKE simulations are encapsulated in Sec.~\ref{sec:qkesim}. The section is composed of four subsections, while we focus only on quasi steady states of radiation fields throughout this paper. We discuss the overall property of flavor conversions in Sec.~\ref{subsec:basicpro}, angular-moment analysis (assessments of closure relations) in Sec.~\ref{subsec:angmom}, and neutrino signal in Sec.~\ref{subsec:signal}. We then summarize the present study in Sec.~\ref{sec:sum}. Unless otherwise noted, we use the metric signature of $- + + +$, and work in the unit with $c = \hbar = 1$, where $c$ and $\hbar$ are the light speed, and the reduced Planck constant, respectively.

\section{Basic equation}\label{sec:baseq}
We solve the mean field quantum kinetic equation for neutrino transport within a three-flavor framework in spherical symmetry,
\begin{equation}
  \begin{split}
& \frac{\partial \barparena{f} }{\partial t} 
+ \frac{1}{r^2} \frac{\partial}{\partial r} ( r^2 \cos \theta_{\nu}  \barparena{f} )
- \frac{1}{ r^2 \sin \theta_{\nu}} \frac{\partial}{\partial \theta_{\nu}} ( \sin^2 \theta_{\nu} \barparena{f} ) \\
& = \barparena{S} - i \xi [\barparena{H},\barparena{f}],
  \end{split}
\label{eq:Flat1DQKE}
\end{equation}
where $t$, $r$, and $\theta_{\nu}$ represent time, radius, and neutrino flight angle (measured from the radial direction) in momentum space, respectively. $f$, $S$, and $H$ denote the density matrix of neutrinos, collision term, and oscillation Hamiltonian, respectively. The upper bar on each variable represents the quantity for antineutrinos.

The neutrino oscillation Hamiltonian is composed of three elements,
\begin{equation}
\barparena{H} = \barparena{H}_{\rm vac} + \barparena{H}_{\rm mat} + \barparena{H}_{\nu \nu}, \label{eq:Hdecompose}
\end{equation}
where
\begin{equation}
\begin{aligned}
&\bar{H}_{\rm vac} = H^{*}_{\rm vac} , \\
&\bar{H}_{\rm mat} = - H^{*}_{\rm mat} ,\\
&\bar{H}_{\nu \nu} = - H^{*}_{\nu \nu}.
\end{aligned}
\label{eq:Hantineutrinos}
\end{equation}
The vacuum potential $H_{\rm vac}$ can be expressed as
\begin{equation}
\begin{aligned}
H_{\rm vac} = \frac{1}{2 E_{\nu}} U
\begin{bmatrix}
        m^{2}_{1} & 0 & 0\\
        0 & m^{2}_{2} & 0 \\
        0 & 0 & m^{2}_{3}
    \end{bmatrix}
 U^{\dagger} ,\\
\end{aligned}
\label{eq:Hvdef}
\end{equation}
where $E_{\nu}$, $U$, and $m_{i(=1,2,3)}$ denote neutrino energy (defined in the laboratory frame), Pontecorvo-Maki-Nakagawa-Sakata (PMNS) matrix, and neutrino mass, respectively. The leading-order term of matter potential ($H_{\rm vac}$) can be written as
\begin{equation}
\begin{aligned}
H_{\rm mat} =
\begin{bmatrix}
        V_e & 0 & 0\\
        0 & 0 & 0 \\
        0 & 0 & 0
    \end{bmatrix}
 ,\\
\end{aligned}
\label{eq:Hmatdef}
\end{equation}
where 
\begin{equation}
V_{e} = \sqrt{2} G_F (n_{e^{-}} - n_{e^{+}}). \label{eq:Vedef}
\end{equation}
In the expression, $n_{e^{\mp}}$ denotes the number density of electron (for minus sign) and positron (for plus sign). $G_F$ represents the Fermi constant. We note that the fluid velocity is assumed to be zero in this study. As a result, the Doppler factor between the laboratory and fluid-rest frames is unity. See \cite{2022PhRvD.106f3011N} for cases with finite fluid velocity in curved spacetimes.

For the sake of simplicity, the matter potential is ignored in this study, while we leave the vacuum potential with reduced mixing angles to take into account refractive effects by medium. We adopt normal mass hierarchy with squared mass differences of $\Delta m^2_{21} = 7.42 \times 10^{-5} {\rm eV^2} $ and $\Delta m^2_{31} = 2.51 \times 10^{-3} {\rm eV^2}$, and we assume that all mixing angles are $10^{-6}$. It should be noted that we adopt different oscillation parameters in the discussion of neutrino signal (Sec.~\ref{subsec:signal}), in which adiabatic Mikheyev–Smirnov–Wolfenstein (MSW) effect is taken into account with experimentally constrained ones. $H_{\nu \nu}$ represents the self-interaction potential under mean field approximation, which can be written as
\begin{equation}
H_{\nu \nu} = \sqrt{2} G_F \int \frac{d^3 q^{\prime}}{(2 \pi)^3} (1 - \sum_{i=1}^{3} \ell^{\prime}_{(i)} \ell_{(i)} ) (f(q^{\prime}) - \bar{f}^{*}(q^{\prime})), \label{eq:Hselfpotedef}
\end{equation}
where $d^3q$ denotes the momentum space volume. $\ell_{i} (i = 1, 2, 3)$ represents the direction of neutrino propagation. The two angles of neutrino flight directions are measured with respect to a radial coordinate basis. By using the polar- ($\theta_{\nu}$) and azimuthal angles ($\phi_{\nu}$) in neutrino momentum space, $\ell_{i} (i = 1, 2, 3)$ can be written as
\begin{equation}
  \begin{split}
&\ell_{(1)} = \cos \theta_{\nu}, \\
&\ell_{(2)} = \sin \theta_{\nu}   \cos \phi_{\nu}, \\
&\ell_{(3)} = \sin \theta_{\nu}   \sin \phi_{\nu}.
  \end{split}
\label{eq:el}
\end{equation}
Since spherically symmetric condition in space assures that neutrino angular distributions are axisymmetric in momentum space ($\phi_{\nu}$-symmetry), we solve Eq.~\ref{eq:Flat1DQKE} with the azimuthal-integrated form.

Following the same prescription in our previous studies \cite{2023PhRvD.107f3033N,2023PhRvL.130u1401N}, the oscillation Hamiltonian is reduced to make the simulations tractable. This is controlled by an attenuation parameter, $\xi$, which is in the right hand side of Eq.~\ref{eq:Flat1DQKE}. We note that both vacuum- and self-interaction terms in the Hamiltonian are attenuated.
In this study, we adopt $\xi=10^{-4}$. 
The same simulation but with $\xi=2 \times 10^{-4}$ under twice higher spatial resolution shows the qualitatively same result \cite{2023PhRvL.130u1401N}, indicating that $\xi=10^{-4}$ can capture the essential trend. Nevertheless, we need to keep in mind that the smallest scale of spatial variations of flavor conversions vary with $\xi$ \cite{2022PhRvL.129z1101N}. We, hence, do not investigate such a detailed structure in this study, but rather focus only on the global property. Another important remark in the present study is that general relativistic effects are neglected. In \cite{2023PhRvL.130u1401N}, we confirm that this simplification does not compromise discussions in the interplay between neutrino transport and flavor conversions, although it should be included in more quantitative arguments.

In the collision term, we consider two charged-current reactions,
\begin{equation}
  \begin{split}  
&e^{-} + p \longleftrightarrow \nu_{e} + n, \\
&e^{+} + n \longleftrightarrow \bar{\nu_{e}} + p,
  \end{split}
\label{eq:ch_reac}
\end{equation}
and two neutral current ones,
\begin{equation}
  \begin{split}  
&\nu + N \longrightarrow \nu + N, \\
&\nu + A \longrightarrow \nu + A,
  \end{split}
\label{eq:neut_reac}
\end{equation}
where $e^{-}$, $e^{+}$, $p$, $n$, $N$, and $A$ denote electron, positron, proton, neutron, nucleon, and heavy nucleus, respectively. $\nu_e$ and $\bar{\nu}_e$ represent electron-type neutrinos and their antipartners, respectively. The charged-current reactions correspond to neutrino emission and absorption processes, and the collision term can be written as
\begin{equation}
\barparena{S}_{ab} =
  \barparena{j}_{a} \delta_{ab}
- \biggl(   \langle \barparena{j} \rangle_{ab}  + \langle \barparena{\kappa} \rangle_{ab} \biggr) \barparena{f}_{ab},
\label{eq:ColQKEEmisAbs}
\end{equation}
where the indices ($a$ and $b$) specify a neutrino flavor. In the expression $j$ and $\kappa$ denote emissivity and absorptivity, respectively. The bracket is defined as
\begin{equation}
\langle Q \rangle_{ab} \equiv \frac{Q_a + Q_b}{2},
\label{eq:defbracket}
\end{equation}
where $Q$ is an arbitrary variable. We adopt $\barparena{j}_{\mu}=\barparena{j}_{\tau}=\barparena{\kappa}_{\mu}=\barparena{\kappa}_{\tau}=0$, since the charged current reactions for heavy leptonic neutrinos ($\nu_x$) can be neglected in the regions considered in this work. 

The neutral current reactions in Eq.~\ref{eq:neut_reac} corresponds to neutrino-nucleon scattering and coherent scatterings by heavy nuclei. Since the rest mass energies of nucleons and heavy nuclei are much higher than the typical energy of neutrinos, these scatterings can be approximately treated as isoenergetic processes. We also neglect all higher order corrections in this reaction, e.g., weak magnetism \cite{2002PhRvD..65d3001H}, strange-quark contributions \cite{2015ApJ...808L..42M} (but see also the argument by \cite{2016PhRvC..93e2801H} regarding the magnitude of nucleon strange helicity) and many body corrections \cite{2006NuPhA.777..356B,2017PhRvC..95b5801H}. The resultant equation can be cast into the following form,
\begin{equation}
  \begin{split}
& \barparena{S}_{ab} (E_{\nu},\Omega) = \\
& - \frac{(E_{\nu})^{2}}{(2 \pi)^3} \int d \Omega^{{\rm \prime}} R (E_{\nu},\Omega,\Omega^{{\rm \prime}}) \times \Bigl( \barparena{f}_{ab} (E_{\nu},\Omega) - \barparena{f}_{ab} (E_{\nu},\Omega^{{\rm \prime}}) \Bigr),
  \end{split}
\label{eq:ColElaSca}
\end{equation}
where 
$\Omega$ and $R$ denote solid angle and scattering kernel, respectively. We note that all thermodynamic quantities relevant to evaluate weak reaction rates are computed with an EOS of \cite{2017JPhG...44i4001F}. We extract these information from matter density ($\rho$), electron-fraction ($Y_e$), and temperature ($T$), which are given from a CCSN model (see Sec.~\ref{sec:model}). For more details of collision term in QKE, we refer readers to \cite{1993PhRvL..70.2363R,1996slfp.book.....R,2000PhRvD..62i3026Y,2014PhRvD..89j5004V,2016PhRvD..94c3009B,2019PhRvD..99l3014R}.

We note a caveat here. We ignore neutrino-electron scatterings and all thermal weak processes of emission and absorption, e.g., nucleon-nucleon bremsstrahlung and electron-positron pair processes. This is because the consistent treatment of these processes requires unfeasible computational resources. Neglecting these reactions would have an influence on neutrino radiation field, in particular for $\nu_x$. Since the impact of flavor conversions hinges on $\nu_x$, this limitation must be kept in mind when interpreting our results.

\begin{figure*}
\begin{minipage}{1.0\textwidth}
    \includegraphics[width=\linewidth]{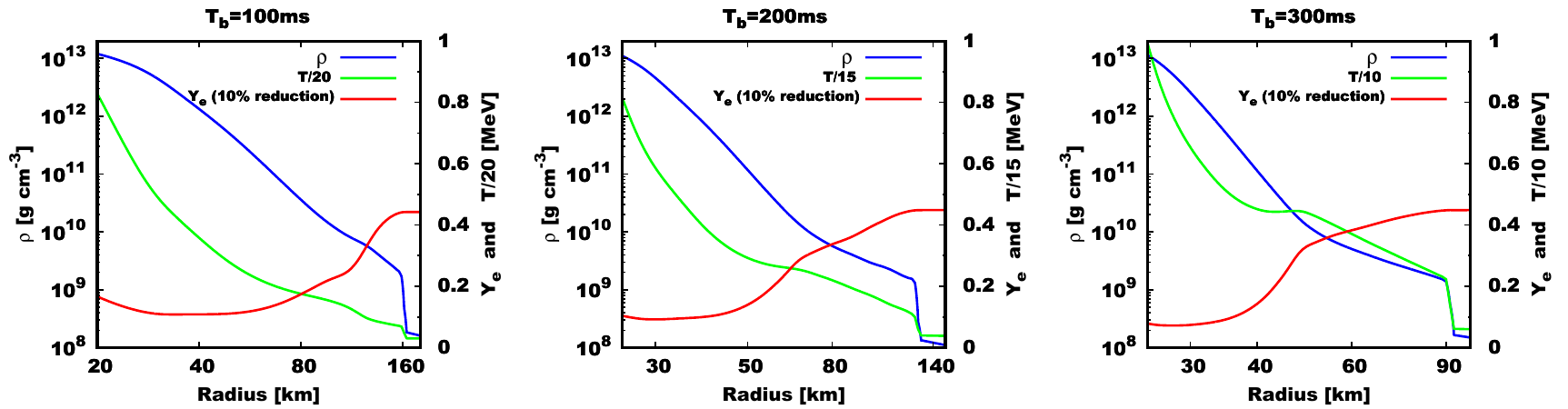}
\end{minipage}
    \caption{Fluid snapshots used in the present study. They are taken from a spherically symmetric CCSN model of 15 solar mass progenitor in \cite{2019ApJS..240...38N}. From left to right, we show fluid profiles for models of ${\rm T}_{\rm b}=100, 200$, and $300$ms, respectively. In each panel, $\rho$, $Y_e$ and $T$ distributions are displayed, and the colors distinguish these quantities. We note that $Y_e$ distribution is reduced by $10\%$ from the original CCSN model; see the text for more details.
}
    \label{graph_fluidprofile_sys}
\end{figure*}

\section{Models}\label{sec:model}
We employ fluid profiles taken from one of our spherically symmetric CCSN models in \cite{2019ApJS..240...38N} (15 solar mass progenitor model in \cite{2002RvMP...74.1015W}). We developed the CCSN model based on our neutrino-radiation hydrodynamic code with full Boltzmann neutrino transport \cite{2014ApJS..214...16N} with taking into account sophisticated neutrino-matter interaction \cite{2019ApJS..240...38N} and with a multi-nuclear EOS based on the variational method for uniform matter, in which nuclear abundance is determined by nuclear statistical equilibrium \cite{2017JPhG...44i4001F}. In this study, we pay special attention to the early post-bounce phase in this model by selecting three different time snapshots: ${\rm T}_{\rm b}=100$, $200$, and $300$ms.

In Fig.~\ref{graph_fluidprofile_sys}, we show radial profiles of $\rho$, $Y_e$ and $T$ in our models. One thing we need to mention here is that we reduce $Y_e$ by $10 \%$ from the original data, which is the same as in \cite{2023PhRvL.130u1401N}. This is motivated by the fact that spherically symmetric CCSN models yield higher $Y_e$ distributions than multi-dimensional models in the vicinity of PNS, which is mainly due to the suppression of convections in spherically symmetric models (see, e.g., \cite{2020MNRAS.492.5764N}). The high $Y_e$ distributions enhance the disparity of number density between $\nu_e$ and $\bar{\nu}_e$, that hampers electron neutrino lepton number (ELN) angular crossings. Since we are currently interested in how flavor conversions change neutrino radiation fields, we deal with this issue pragmatically by reducing $Y_e$. We checked the sensitivity of our results to the degree of $Y_e$-reduction by carrying out another simulation with $7 \%$ reduction for ${\rm T}_{\rm b}=300$ ms snapshot, and we confirmed that the qualitative trend did not change. We note, however, that more detailed modelings, i.e., three-dimensional (3D) quantum kinetic neutrino transport simulations based on fluid profiles from 3D CCSN models will be necessary to quantify the impacts of neutrino oscillations on CCSN dynamics accurately, in which these models naturally generate ELN crossings without changing $Y_e$; these studies have to be deferred until more computational resources are available, though.

We employ our GRQKNT code to solve Eq.~\ref{eq:Flat1DQKE}. The methodology and implementation of our code are described in \cite{2022PhRvD.106f3011N}. Before we run quantum kinetic simulations, classical simulations ($\barparena{H}$ is set to be zero in Eq.~\ref{eq:Flat1DQKE}) are performed until the radiation field settles into a steady state. The obtained neutrino distributions are compared with results of quantum kinetic simulations, and also used for their initial condition. For time-efficient simulations, the computational domain varies among different fluid profiles: $20$km-$180$km, $25$km-$150$km, and $20$km-$100$km for the time snapshot of ${\rm T}_{\rm b}=100,$ $200$, and $300$ms, respectively. The location of each inner boundary is determined by matter density of $\rho \sim 10^{13} {\rm g/cm^3}$ (see also Fig.~\ref{graph_fluidprofile_sys}), where all flavors of neutrinos are almost in thermal and chemical equilibrium with matter. Such equilibrium states are compatible with a Dirichlet boundary condition at the inner boundary, in which outgoing neutrinos are assumed to be in purely flavor eigenstates and Fermi-Dirac distributions\footnote{For incoming neutrinos, we impose a free-streaming condition for all elements of density matrix of neutrinos.}. The chemical potential of $\nu_e$ is computed by $\mu_p - \mu_n + \mu_e$ ($\mu_{i}$ denotes the chemical potential of $i$-particle) and we adopt its negative value for $\bar{\nu}_e$. For all heavy leptonic neutrinos, the chemical potential is assumed to be zero. Given the chemical potential and matter temperature, we determine Fermi-Dirac distributions for each species of neutrinos. The outer boundary is, on the other hand, determined so as to cover the post-shock region, and hence it is set ahead of a stalled shock wave. We impose the free-streaming boundary condition for outgoing neutrinos, while no incoming neutrinos are injected from the outer boundary.

We cover the computational domain with a non-uniform radial grid with $12,288$ points. The grid width increases with radius logarithmically, while the innermost resolution is set to be $30$cm. This spatial resolution has the ability to capture a wavelength of "attenuated" flavor conversions by $\gtrsim 10$ grid points (if FFCs happen). We employ 12 grid points for the neutrino energy in the region of $0$MeV $\le E_{\nu} \le 80$MeV. It is a logarithmically uniform grid while the lowest grid covers from $0$MeV to $2$MeV. The neutrino angular grid is uniform with $96$ points, that covers from $-1 \le \cos{\theta_{\nu}} \le 1$. Each simulation is run until $t=1$ms, that is long enough for the neutrinos to reach a quasi-steady state. Similar as our previous paper \cite{2023PhRvL.130u1401N}, we limit ourselves to considering the time-averaged properties of neutrino radiation field in the quasi-steady state, since the temporal feature is sensitive to the attenuation parameter \cite{2022PhRvL.129z1101N}. To compute the time-averaged quantities, we extend each simulation up to $1.05$ms and then compute the time averaged quantities during the interval of $1 {\rm ms} \le t \le 1.05 {\rm ms}$.

\section{Linear stability analysis}\label{sec:linana_ini}

We carry out a linear stability analysis for steady-state neutrino radiation fields obtained from classical neutrino transport, which correspond to initial conditions for quantum kinetic simulations. As a remark, we work in the metric signature of $- + + +$ throughout this paper; hence the resultant linearized equation is different from that with $+ - - -$ which is frequently used in literature. This is just a matter of convention and the result of dispersion relation is unchanged.

It should be mentioned that we simplify our stability analysis by using the following approximations. First, we carry out this stability analysis under two-flavor framework. This is a reasonable approach, since we assume $\nu_{\mu}=\nu_{\tau}$ in classical neutrino transport (but see \cite{2020JCAP...01..005C,2021PhRvD.103f3013C} for cases with distinguishing $\nu_{\mu}$ and $\nu_{\tau}$). The resultant linearized equation for the off-diagonal term of neutrino density matrix can be written as\footnote{This is a local stability analysis; thus the QKE is written in a local orthonormal basis.},
\begin{equation}
    \begin{split}
    i(\partial_t&+\boldsymbol{v}\cdot\nabla) f_{ex} = i S_{ex} -\xi\omega_{\mathrm{V}}f_{ex} \\
    &- \sqrt{2}G_{\mathrm{F}}\xi f_{ex} v^{\mu}\int\frac{d^3 q^{\prime}}{(2\pi)^3} v_{\mu}^{\prime}(f_{ee}(q^{\prime})-f_{xx}(q^{\prime})) \\
    &+ \sqrt{2}G_{\mathrm{F}}\xi (f_{ee}-f_{xx}) v^{\mu}\int\frac{d^3 q^{\prime}}{(2\pi)^3} v_{\mu}^{\prime}f_{ex}(q^{\prime}),
    \end{split}
\end{equation}
where $v^{\mu}=(1,\boldsymbol{v})$, $\omega_{\mathrm{V}}$ is a corresponding vacuum frequency, while we follow the flavor isospin convention as $\overline{f}(E_{\nu})\equiv -f(-E_{\nu})$ to denote antineutrinos \cite{2018JCAP...12..019A}.
We adopt the plane wave ansatz to the off-diagonal part 
\begin{equation}
f_{ex} \propto \tilde{Q}\, e^{i K^{\mu}x_{\mu}} = \tilde{Q}\, e^{-i(\Omega t-\boldsymbol{K}\cdot\boldsymbol{x})},
\end{equation}
and then recast the equation into
\begin{equation}
    \begin{split}
    -v^{\mu}k_{\mu}\tilde{Q} &= i\tilde{S}_{ex} - \xi \omega_{\mathrm{V}} \tilde{Q} \\
    &+ \sqrt{2}G_{\mathrm{F}}\xi (f_{ee}-f_{xx}) v^{\mu}\int\frac{d^3 q^{\prime}}{(2\pi)^3} v_{\mu}^{\prime}\tilde{Q}(q^{\prime}),
    \end{split}
\end{equation}
where $k_{\mu} \equiv (\omega,\boldsymbol{k}) = K_{\mu}-\xi\Phi_{\mu}$ with $\Phi_{\mu} \equiv \sqrt{2}G_{\mathrm{F}}\int\frac{d^3 q^{\prime}}{(2\pi)^3} v_{\mu}^{\prime}(f_{ee}(q^{\prime})-f_{xx}(q^{\prime}))$.
Note that $k=0$ is a zero mode in a rotating frame, not always identical to a true homogeneous mode $K=0$ unless $\Phi_z=0$ along the spatial direction.

\begin{figure*}
\begin{minipage}{1.0\textwidth}
    \includegraphics[width=\linewidth]{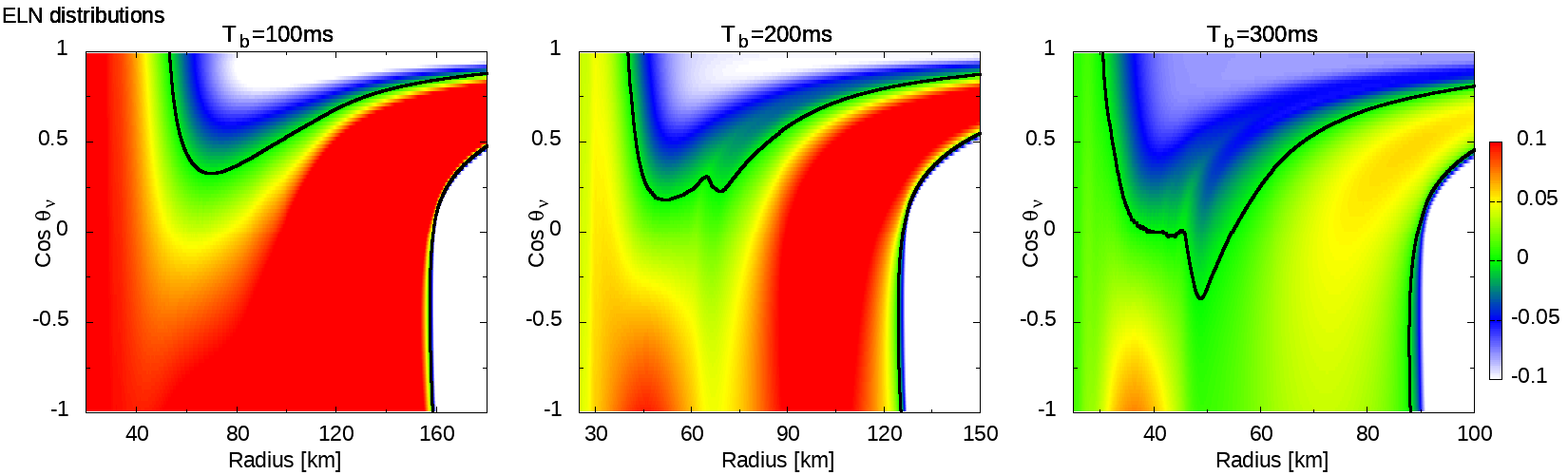}
\end{minipage}
    \caption{Color map of ELN distributions, $(f_{ee}-\bar{f}_{ee})$, normalized by the sum of all flavors of neutrinos and antineutrinos, as functions of radius and neutrino flight angle. The distributions are taken from classical neutrino transport models ($H=\bar{H}=0$) at the end of each simulation (neutrinos are in quasi-steady states). From left to right, we show results for models of ${\rm T}_{\rm b}=100, 200,$ and $300$ms, respectively. In each panel, we also highlight $(f_{ee}=\bar{f}_{ee})$ by black solid lines.
}
    \label{graph_Radi_vs_angular_iniELN}
\end{figure*}

We adopt another simplification in collision term. Although we include neutral current reactions in our quantum kinetic simulations, we neglect them in this stability analysis. One of the reasons of this simplification is that these reactions involve integrations over neutrino momentum space. This makes the analysis far complex and also computationally more expensive. Despite of the complexity and requiring extensive computations, these neutral current reactions would play a subdominant role compared to charged current ones in the present study; the reason is as follows. As mentioned already, the neutrino-matter interactions are important for driving collisional instability. The growth rate of the instability is proportional to neutrino-matter reaction rate ($R$) or to $\sqrt{R n_{\nu}}$ (where $n_{\nu}$ denotes the neutrino number density) in the resonance region \cite{2022arXiv221203750X,2023arXiv230206263L}. This suggests that the instability can develop vigorously only in optically thick region. On the other hand, the neutrino angular distribution is almost isotropic in the region, implying that the detailed balance between in- and out-scatterings is nearly satisfied. For this reason, the collisional instability in the optically thick region would be less affected by scatterings at least in the linear regime. One thing we do notice here is that emission and absorption processes are qualitatively different from scatterings, since they can trigger collisional instability even when neutrino are in thermal and chemical equilibrium with matter, i.e., isotropic distributions.

\begin{figure*}
\begin{minipage}{1.0\textwidth}
    \includegraphics[width=\linewidth]{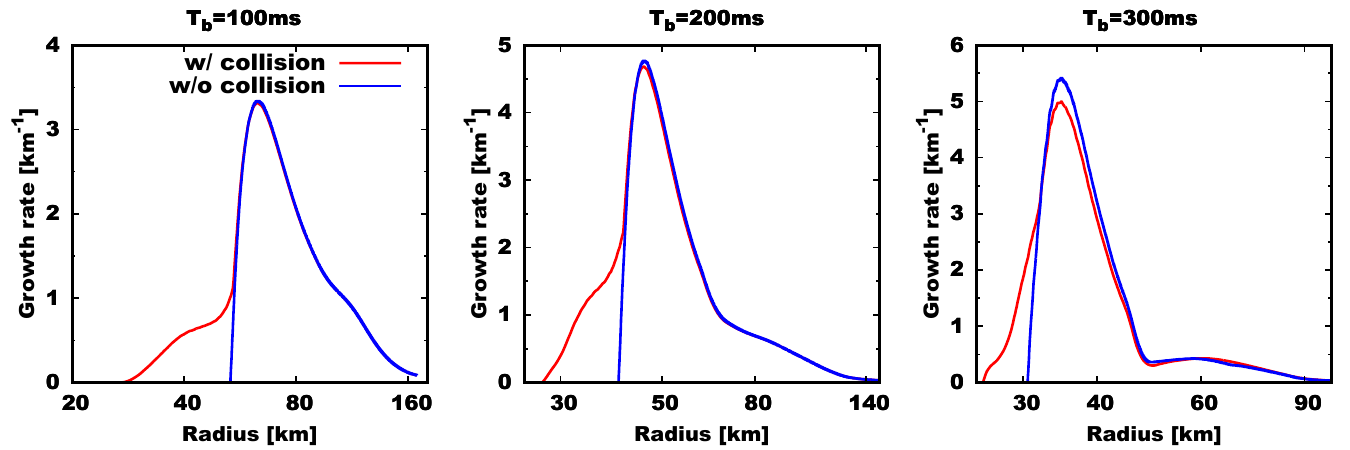}
\end{minipage}
    \caption{Radial profiles of the maximum growth rate of flavor conversion as a function of radius. From left to right, we show the result for models of ${\rm T}_{\rm b}=100, 200$, and $300$ms, respectively. In this analysis, we take into account the attenuation parameter ($\xi=10^{-4}$). Red and blue lines show the result with and without collision terms, respectively, which is useful to see which collisional instability or FFC dominates flavor conversions; see the text for more details.
}
    \label{graph_Disp_ini}
\end{figure*}

\begin{figure*}
\begin{minipage}{1.0\textwidth}
    \includegraphics[width=\linewidth]{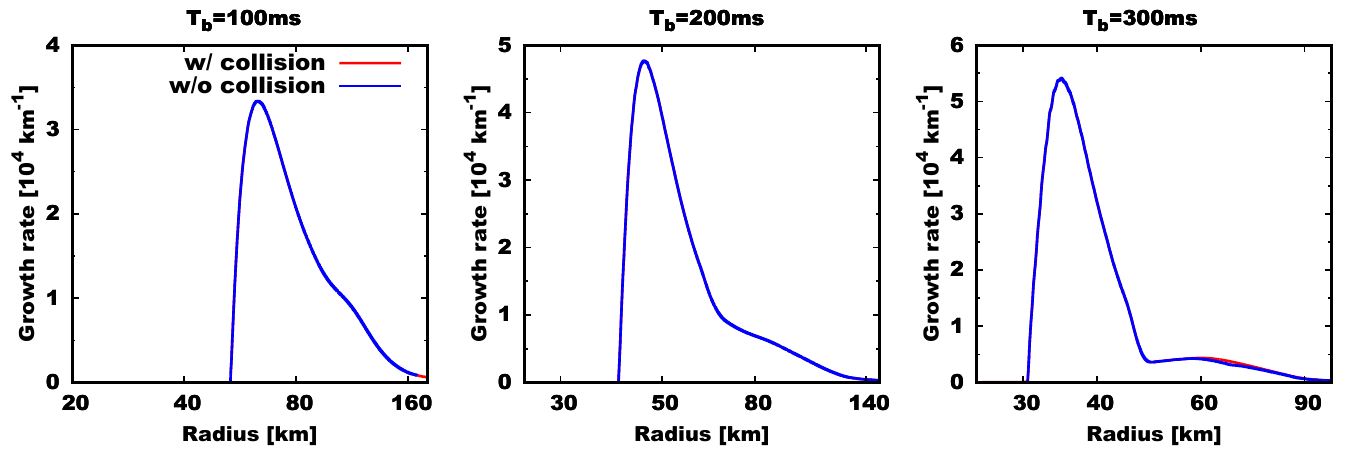}
\end{minipage}
    \caption{The same figure as Fig.~\ref{graph_Disp_ini} but for cases with no attenuation of oscillation Hamiltonian, i.e., $\xi=1$ .
}
    \label{graph_Disp_ini_NoAttenu}
\end{figure*}

For the emission and absorption process, the off-diagonal component can be cast into the following form,
\begin{equation}
    \begin{split}
    \tilde{S}_{ex} &= -\left[\langle j \rangle_{ex} +\langle \kappa \rangle_{ex} \right]\tilde{Q} \\
    &\equiv -R_{\mathrm{EA}}\tilde{Q},
    \end{split}
\end{equation}
and see Eq.~\ref{eq:defbracket} for the notation.
Then, nontrivial solutions for $\tilde{Q}$ are given by
\begin{equation}
    \mathrm{det}\left[\Pi^{\mu\nu}(k)\right] = 0,
\end{equation}
where
\begin{equation}
    \Pi^{\mu\nu} = \eta^{\mu\nu}+\sqrt{2}G_{\mathrm{F}}\int\frac{d^3 q}{(2\pi)^3}\xi(f_{ee}-f_{xx})\frac{v^{\mu}v^{\nu}}{v\cdot k -\xi\omega_{\mathrm{V}} -i R_{\mathrm{EA}}}.
\end{equation}
If there is a positive $\mathrm{Im}\,\omega$ in the dispersion relation, neutrinos undergo a flavor instability and can exponentially grow in time. As an important remark, our stability analysis does not impose a homogeneous condition. This is necessary to prevent us from overlooking the most unstable wave number.

\begin{figure*}
\begin{minipage}{1.0\textwidth}
    \includegraphics[width=\linewidth]{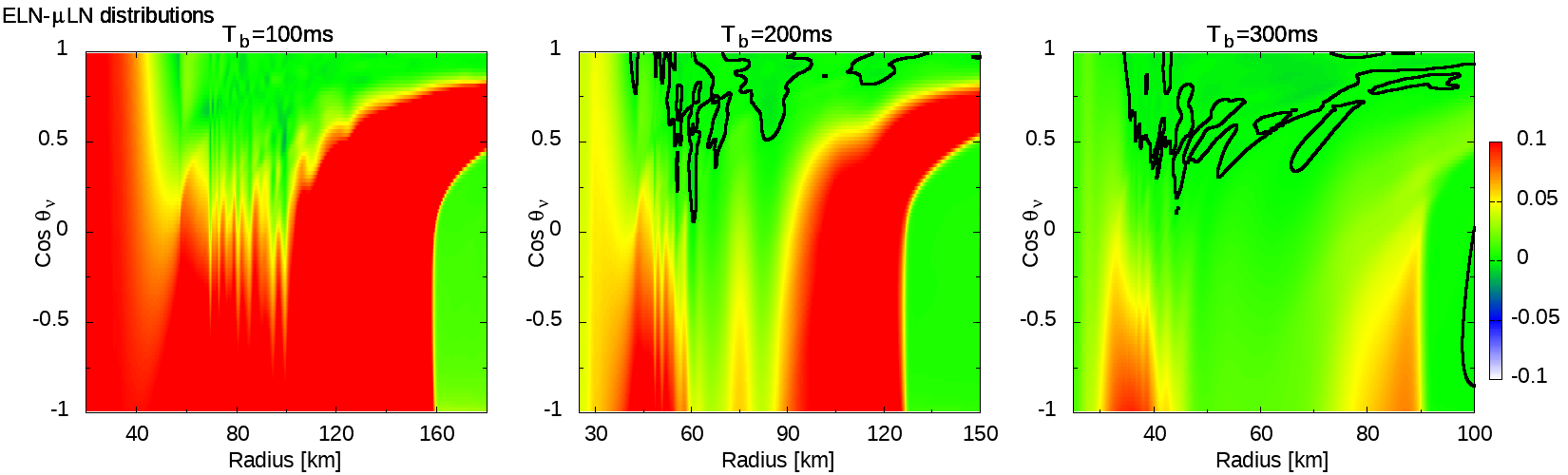}
\end{minipage}
    \caption{Same figure as Fig.~\ref{graph_Radi_vs_angular_iniELN} but for the time-averaged ELN-$\mu$LN distributions in quasi-steady states, obtained from quantum kinetic neutrino transport simulations.
}
    \label{graph_Radi_vs_angular_FFCtave_ELNmuLN}
\end{figure*}

Before we discuss the result of linear stability analysis, we show ELN distributions as functions of radius and neutrino angle at the end of classical simulations  in Fig.~\ref{graph_Radi_vs_angular_iniELN}. As displayed in the figure, there are ELN angular crossings in all models, indicating that FFC should occur if we turn on the oscillation Hamiltonian. There are two remarks about the property of ELN crossings. First, the similar crossings in the post-shock region are also observed in recent multi-dimensional CCSN models \cite{2019ApJ...886..139N,2021PhRvD.104h3025N}, and they belong to the so-called Type-II crossing in \cite{2021PhRvD.104h3025N}. As discussed in these studies, this type of crossing is generated by strong asymmetric neutrino emission, in which $\bar{\nu}_e$ emission can overwhelms $\nu_e$ in the low $Y_e$ environment. Such asymmetric $Y_e$ distributions near the PNS surface can be triggered by lepton-emission self-sustained asymmetry (LESA) \cite{2014ApJ...792...96T,2019PhRvD.100f3018W,2019ApJ...881...36G,2019MNRAS.487.1178P} or coherent lateral motions of matter associated with PNS kick \cite{2019ApJ...880L..28N}. Although our CCSN models are in spherically symmetry, the reduction of $Y_e$ leads to the qualitatively similar matter profiles as in these multi-dimensional models. Second, $\bar{\nu}_e$ always dominates over $\nu_e$ for incoming neutrinos in the pre-shock region. This is mainly due to coherent scatterings of neutrinos by heavy nuclei, as discussed in \cite{2020PhRvR...2a2046M}. It should be mentioned, however, that the crossing depth is tiny and the expected growth time scale of FFC is a few orders of magnitudes smaller compared to that in the post-shock region. This indicates that attenuation of neutrino Hamiltonian would suppress the development of FFCs in this region (see \cite{2021PhRvD.104h3035Z,2022JCAP...03..051A} for FFCs in cases with no attenuation of Hamiltonian). As shown later, this type of crossings in ELN-XLN angular distributions disappears in asymptotic states obtained quantum kinetic simulations. The detail shall be discussed in Sec.~\ref{sec:qkesim}.

Figure~\ref{graph_Disp_ini} portrays radial profiles of the maximum growth rate of flavor conversion at each spatial position. We note that the growth rate is computed from the linearlized QKE with $\xi=10^{-4}$ attenuation parameter. In the same figure, we also show the same quantity but computed without collision term (red colors). This result suggests that the dominant mode of flavor conversion varies with radius. In the inner region (or optically thick region) where no ELN angular crossings are involved, flavor conversions are triggered by collisional instability. In fact, these regions are stable if we ignore collision terms. However, the growth rates between with and without collision terms overlap each other at large radii, indicating that FFC becomes dominant in the region where ELN crossings appear.

It is also interesting to see results of linear stability analysis with no attenuation of Hamiltonian, which are displayed in Fig.~\ref{graph_Disp_ini_NoAttenu}. Compared to Fig.~\ref{graph_Disp_ini}, we confirm that the growth rate of FFCs becomes much larger than that of collisional instability. This result is consistent with the argument that the growth rate of FFC is simply proportional to $\xi^{-1}$, whereas that of collisional instability is scaled by $\xi^{-0.5}$ in the resonance region \cite{2022arXiv221203750X,2023arXiv230206263L} otherwise it does not depend on $\xi$.

Before closing this section, one thing we should remark is that the growth rate does not represent how vigorous flavor conversions occur in non-linear phase. In fact, as discussed in our previous paper \cite{2023PhRvD.107j3022Z}, the degree of flavor conversion in non-linear regime depends on structures of neutrino distributions. In the next section, we delve into the non-linear phase of flavor conversions.

\section{QKE simulations}\label{sec:qkesim}
In this section, we present results of numerical simulations of quantum kinetic neutrino transport. We start with discussing overall properties of flavor conversions in Sec.~\ref{subsec:basicpro}. We then quantify Eddington tensors for diagonal elements of density matrix to assess classical closure relations in Sec.~\ref{subsec:angmom}. In Sec.~\ref{subsec:signal}, we study the neutrino signal in three representing neutrino detectors: Super-Kamiokande (SK) \cite{2016APh....81...39A}, the deep underground neutrino experiment (DUNE) \cite{2021arXiv210313910A}, and the Jiangmen Underground Neutrino Observatory (JUNO) \cite{2022PrPNP.12303927J}. We then discuss how flavor conversions leave observable imprints in neutrino signal.

\subsection{Basic properties}\label{subsec:basicpro}

\begin{figure*}
\begin{minipage}{1.0\textwidth}
    \includegraphics[width=\linewidth]{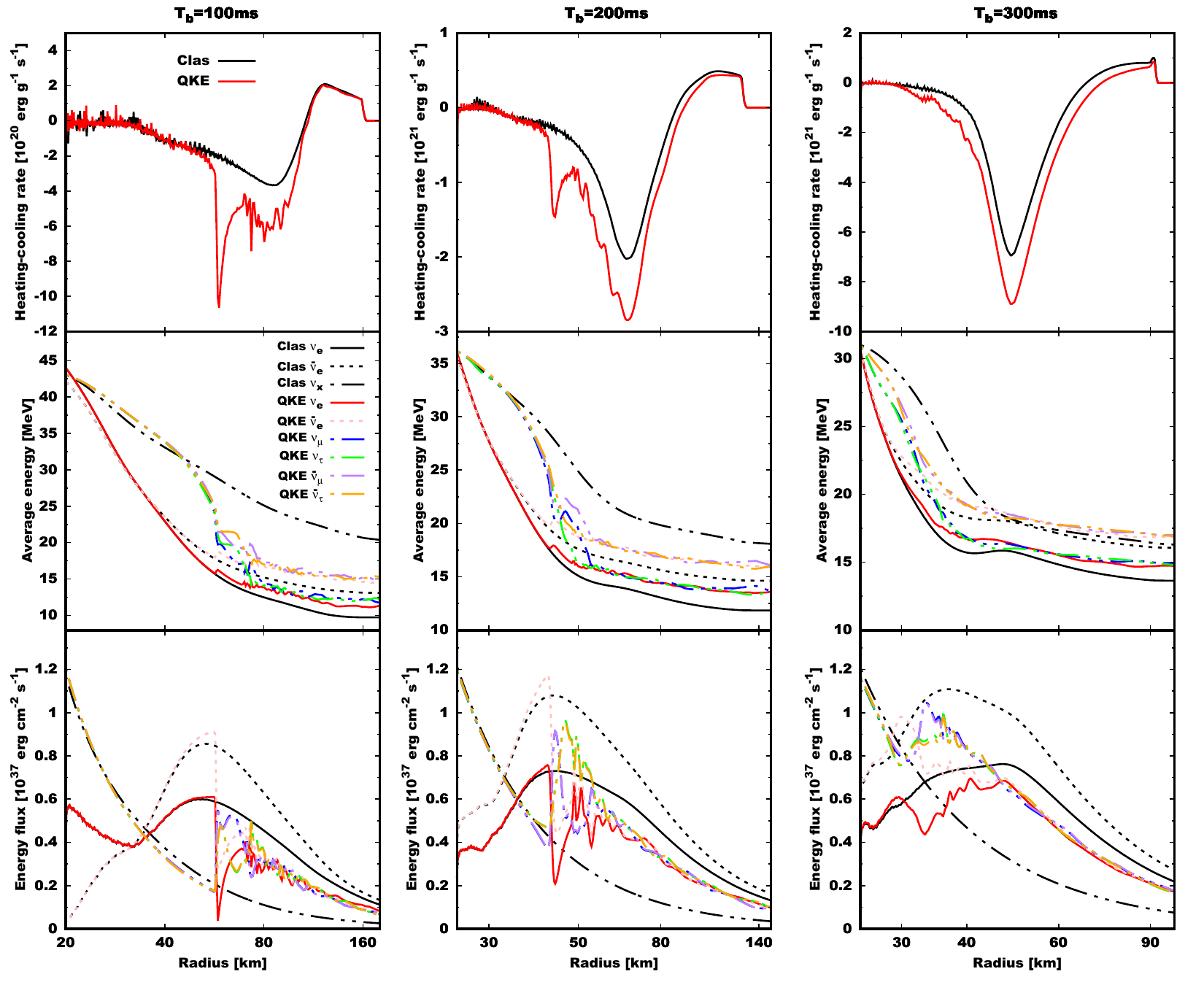}
\end{minipage}
    \caption{Radial profiles of some important time-averaged quantities in quantum kinetic neutrino transport simulations. From top to bottom, we show the result of neutrino heating-cooling rate, average energy, and energy flux of all flavors of neutrinos. As a reference, we display their counterparts in cases with classical neutrino transport as black colors (denoted by "Clas"), while the line type distinguishes neutrino flavors. Other colored lines correspond to results of quantum kinetic simulations. From left to right, we show the results for models of ${\rm T}_{\rm b}=100, 200$, and $300$ms, respectively.
}
    \label{graph_HeatCool_Aveene_Eneflux}
\end{figure*}

Our previous works of global simulations \cite{2022PhRvL.129z1101N,2023PhRvD.107f3033N} show that FFCs evolve towards disappearance of ELN-XLN angular crossings (XLN denotes heavy neutrinos lepton number). This argument is supported by a stability analysis in non-linear regime \cite{2023PhRvD.107j3022Z} and by many local simulations including other groups (see, e.g., \cite{2023arXiv230711129X}). However, we have discusses so far these asymptotic states of FFCs with idealized simulations, in which we purposely neglect neutrino-matter interactions to focus only on properties of FFCs. It is interesting to check if the disappearance of the ELN-XLN crossing holds in cases with neutrino-matter interactions.

In Fig.~\ref{graph_Radi_vs_angular_FFCtave_ELNmuLN}, we display the similar quantity as Fig.~\ref{graph_Radi_vs_angular_iniELN} but for the time-averaged ELN-$\mu$LN ($\mu$LN denotes $\mu$ neutrinos lepton number) distributions for the time-averaged quantities in quasi-steady states obtained from quantum kinetic simulations. Regardless of models, the crossings almost disappear, indicating that the disappearance of crossing is an intrinsic characteristic of FFCs even with neutrino-matter interactions. On the other hand, we find a qualitatively different results from those in our previous studies \cite{2022PhRvL.129z1101N,2023PhRvD.107f3033N}. ELN-$\mu$LN becomes almost positive in the entire computational domain, whereas previous studies suggested that it is usually negative in cases with $\bar{f}_{\nu_e} > f_{\nu_e}$ at $\cos \theta_{\nu} \sim 1$ (see, e.g., the right panel of Fig.~14 in \cite{2023PhRvD.107f3033N}). This indicates that neutrino-matter interactions play an important role on determining asymptotic states of flavor conversions.

In cases with no neutrino-matter interactions, asymptotic states of FFCs can be characterized by conserved quantities, e.g., neutrino number density or flux, which depends on boundary conditions \cite{2023PhRvD.107j3022Z,2023PhRvD.107l3021Z}. This argument is also supported by the results of our previous global simulations in \cite{2022PhRvL.129z1101N,2023PhRvD.107f3033N}. However, there are no conserved quantities in cases with neutrino-matter interactions, which can naturally lead to different asymptotic states. It is very interesting to analyze what quantity (instead of conserved quantities) is a key to characterize the FFC and how it changes by global advection. Addressing the issue, however, requires detailed studies, which is deferred to our future work.

It is also worth to note that ELN-$\mu$LN becomes positive for incoming neutrinos in the pre-shock region, which is qualitatively different from the initial distribution\footnote{We note that $\mu$LN is zero at the initial condition, indicating that ELN distributions are equal to ELN-$\mu$LN ones.}. One may wonder if the disappearance of the crossing is due to FFCs triggered by scatterings of heavy nuclei \cite{2020PhRvR...2a2046M,2021PhRvD.104h3035Z}. As already pointed out in Sec.~\ref{sec:linana_ini}, the neutrino number for incoming neutrinos (which triggers FFCs) is small in the pre-shock region, and thus the attenuation of neutrino oscillation Hamiltonian weaken FFCs substantially. This suggests that the disappearance of crossing is not due to FFCs, and there should be another mechanism. We will revisit this issue later.

\begin{figure*}
\begin{minipage}{1.0\textwidth}
    \includegraphics[width=\linewidth]{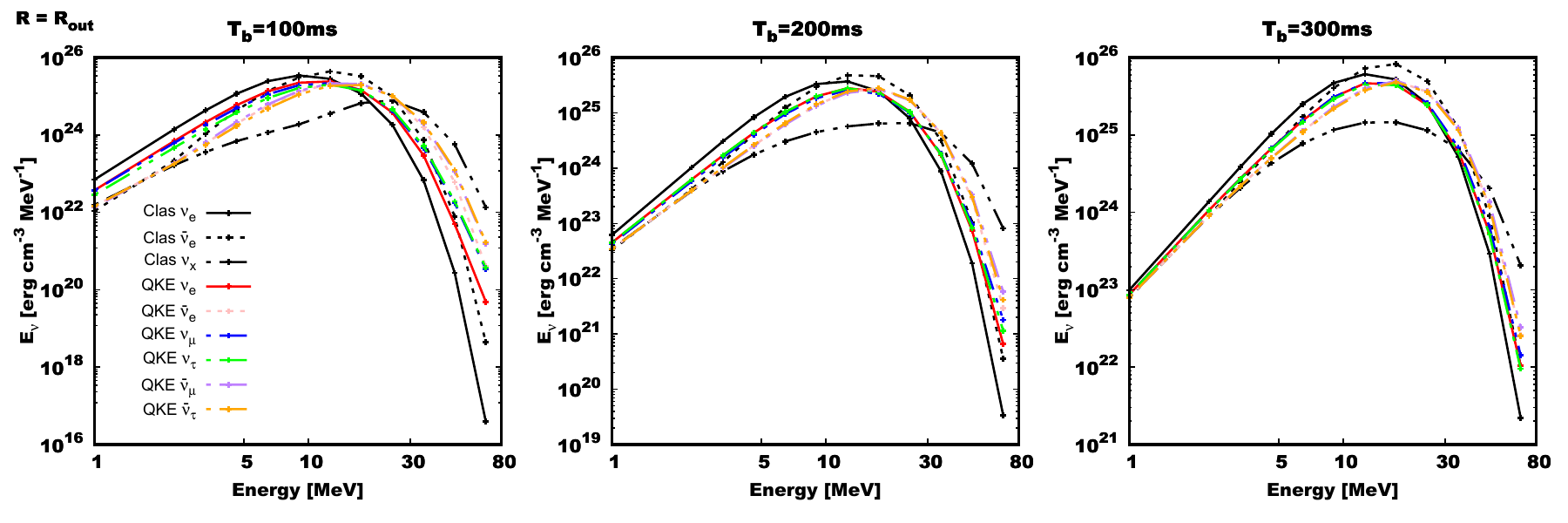}
\end{minipage}
    \caption{Time-averaged energy spectra for neutrino number flux measured at the outer boundary. The line colors and types are the same as those used in Fig.~\ref{graph_HeatCool_Aveene_Eneflux}. From left to right, we show the results for models of ${\rm T}_{\rm b}=100, 200$, and $300$ms, respectively.
}
    \label{graph_NumfluxSpect}
\end{figure*}

We now turn our attention to discussions of some key quantities in neutrino radiation field to consider feedback to hydrodynamics. The upper panels in Fig.~\ref{graph_HeatCool_Aveene_Eneflux} show the net gain or loss of energy rate by neutrino emission and absorption through charged current reactions (see Eq.~\ref{eq:ch_reac}). Although the detailed features depend on models, neutrino cooling is commonly enhanced in the optically thick region, while the heating rate in the gain region is suppressed by flavor conversions. As shown in the middle and bottom panels, the overall trend of the impact of flavor conversions on the average energy and energy flux of neutrinos is similar among all models. Flavor conversions increase the average energy of $\nu_e$ and $\bar{\nu}_e$ due to the mixing of $\nu_x$, whereas their energy fluxes are reduced. Our result suggests that the latter effect dominates over the former, leading to the reduction of neutrino heating in the gain region. On the other hand, the increase of $\nu_x$ energy flux accounts for the enhancement of neutrino cooling in the optically thick region. These overall trends are consistent with those found in our previous study \cite{2023PhRvL.130u1401N}.

We note that flavor equipartitions are nearly achieved but there remain remarkable differences between neutrinos and antineutrinos. This feature can be clearly seen in the average energy of neutrinos at large radii (the middle panels of Fig.~\ref{graph_HeatCool_Aveene_Eneflux}). The average energy of neutrinos are almost identical among all flavors, and the trend is the same for antineutrinos. However, the antineutrinos have the systematically higher energy than neutrinos at large radii. This is due to the fact that low $Y_e$ (or neutron rich) environment makes $\bar{\nu}_e$ to decouple with matter at the inner region than $\nu_e$. Since matter temperature increases with decreasing radius, the average energy of $\bar{\nu}_e$ is higher than $\nu_e$. This is a well-known characteristic of neutrino radiation field in CCSNe, while our results suggest that the trend holds even when flavor conversions occur. We conclude that the difference of neutrino-matter interactions between $\nu_e$ and $\bar{\nu}_e$ mainly accounts for the difference of flavor equipartitions between $\nu$ and $\bar{\nu}$ in cases with occurrences of strong flavor conversions in post-shock regions.

This result would be an important information in the study of neutrino signal from CCSNe. In fact, we usually assume that $\nu_x$ and $\bar{\nu}_x$ have almost identical properties each other in the conventional CCSN theory, and therefore the independent number of neutrino species is essentially three: $\nu_e$, $\bar{\nu}_e$, and $\nu_x$. However, our result suggests that the independent signal becomes only two in cases with strong flavor conversions: $\nu$ and $\bar{\nu}$. We shall discuss in Sec.~\ref{subsec:signal} how this property can be used in data analysis of neutrino observations. Our result also suggests that we can disentangle the effect of on-shell muon weak interactions from flavor conversions in neutrino signal. As discussed in previous works \cite{2017PhRvL.119x2702B,2020PhRvD.102l3001F,2020PhRvD.102b3037G}, the muon reactions breaks the degeneracy of $\nu_{\mu}$ and $\nu_{\tau}$ (and their antipartners) in neutrino signal, while flavor conversions make neutrinos or antineutrinos to be flavor equipartitions\footnote{One thing we do notice is that the neutrino mixing angles in vacuum oscillation term may generate the difference between $\nu_{\mu}$ and $\nu_{\tau}$, but the effect is negligibly small in the case with FFCs. This suggests that the vacuum term does not compromise the present study.}. Another intriguing question associated with this discussion is that the on-shell muon appearance and their interactions with neutrinos may give an impact on flavor conversions (see also \cite{2021PhRvD.103f3013C}). We do not know at the moment how it changes the dynamics and asymptotic state of flavor conversions. This issue needs to be addressed with appropriate treatments of muon weak interactions, which is, however, beyond the scope of this paper.

Here, we revisit a discussion of why ELN-XLN crossings in our quantum kinetic simulations disappear in the pre-shock region (see Fig.~\ref{graph_Radi_vs_angular_FFCtave_ELNmuLN}). 
In cases with classical models, XLN is assumed to be zero, and the ELN-XLN (or ELN) crossing appears due to the disparity of neutrino flux and its average energy between $\nu_e$ and $\bar{\nu}_e$. More specifically, the higher average energy of $\bar{\nu}_e$ than $\nu_e$ facilitates coherent scatterings by heavy nuclei, and therefore $\bar{\nu}_e$ can dominate over $\nu_e$ in the radially incoming neutrinos, that can trigger ELN-crossing (see \cite{2020PhRvR...2a2046M} for more details). In quantum kinetic models, however, heavy-leptonic neutrinos and antineutrinos are no longer identical to each other (i.e., $\nu_x \neq \bar{\nu}_x$), and flavor equipartitions are nearly achieved for out-going neutrinos and antineutrinos, i.e., $\nu_e \sim \nu_x$ and $\bar{\nu}_e \sim \bar{\nu}_x$. This implies that back-scattered neutrinos by coherent scatterings with heavy nuclei are nearly equal between $\nu_e$ and $\nu_x$ (and likewise for $\bar{\nu}_e$ and $\bar{\nu}_x$). As a result, the ELN-XLN is zero for not only outgoing- but also incoming neutrinos, that results in suppressing ELN-XLN crossings. Our result suggests that FFCs in pre-shock region would be suppressed if outgoing neutrinos undergo strong flavor conversions in post-shock regions. This trend would hold in cases with no attenuation of neutrino oscillation Hamiltonian, since the flavor equipartition in outgoing neutrinos would be achieved there.


In Fig.~\ref{graph_HeatCool_Aveene_Eneflux}, we show that the overall trend is essentially the same among all models, but it is worthy to discuss some features depending on models. First, the difference of neutrino heating in the gain region between classical- and quantum simulations is subtle for ${\rm T}_{\rm b}=100$ms, and it looks that the difference increases with ${\rm T}_{\rm b}$. It should be noted, however, that our results may not be enough to draw a robust conclusion on this trend. As mentioned above, the impact of FFCs on net neutrino heating rate is determined by a delicate balance between change of average energy and energy flux, while such a quantitative discussion requires accurate modelings of $\nu_x$ radiation field. However, our simplifications in $\nu_x$-matter interactions have an influence on $\nu_x$ properties. We also note that the degree of $Y_e$ reduction affect this quantitative argument. These limitations in the present study should be borne in mind. 

Next, we find that the sharp increase of neutrino cooling in the optically thick region for models of ${\rm T}_{\rm b}=100$ and $200$ms (see at $r \sim 60$km and $40$km for the former and the latter model, respectively, in Fig.~\ref{graph_HeatCool_Aveene_Eneflux}), whereas such a rapid change is not observed in the model of ${\rm T}_{\rm b}=300$ms. This model-dependent feature can be understood as follows. For models of ${\rm T}_{\rm b}=100$ and $200$ms, the $\nu_e$ ($\bar{\nu}_e$) number density is remarkably higher than $\nu_x$ ($\bar{\nu}_x$) at the innermost region where ELN crossings occur (hereafter we denote the radius as $r_{\rm IFC}$). As a result, both $\nu_e$ and $\bar{\nu}_e$ densities (and also fluxes) are substantially reduced by FFCs, causing the sharp increase of neutrino cooling. On the other hand, the difference of neutrino number density among different species of $\bar{\nu}$ at $r_{\rm IFC}$ is moderate for the model of ${\rm T}_{\rm b}=300$ms, leading to relatively smooth transition of neutrino radiation field (see right panels of Fig.~\ref{graph_HeatCool_Aveene_Eneflux}), and therefore there is no sharp transition of neutrino cooling. This argument suggests that the smooth transition would not be affected by attenuation of Hamiltonian, since the degree of neutrino flavor conversion is determined by the disparity between $\nu_e$ and $\nu_x$ or $\bar{\nu}_e$ and $\bar{\nu}_x$.

As discussed above, our result in models of ${\rm T}_{\rm b}=100$ and $200$ms suggests that neutrino radiation field can be changed substantially in the very small spatial scale (see again at $r \sim 60$km and $40$km for the former and the latter model, respectively, in Fig.~\ref{graph_HeatCool_Aveene_Eneflux}) in cases that there are large deviations between $\nu_e$ ($\bar{\nu}_e$) and $\nu_x$ ($\bar{\nu}_x$) before flavor conversion happen. This implies that high spatial resolutions are indispensable for accurate modelings of quantum kinetic neutrino transport. We also note that the transition scale in reality is much smaller than that obtained in the present study, suggesting that attenuating Hamiltonian makes these global simulations tractable.

As mentioned already, the flavor equipartition is nearly achieved in our simulations. However, flavor-dependent features still remain in particular at high energies. As shown in Fig.~\ref{graph_NumfluxSpect}, which displays the energy spectrum of neutrino number flux measured at the outer boundary, heavy-leptonic neutrinos are systematically higher than electron-type ones in the energy of $E_{\nu} \gtrsim 30$MeV. This is indicative of the fact that, for some neutrinos, flavor conversions subside inside of energy spheres of $\nu_e$ and $\bar{\nu}_e$. We note that the location of these spheres increases with neutrino energies, indicating that the high energy $\nu_e$ and $\bar{\nu}_e$ undergo strong absorptions before escaping from their energy spheres. On the contrary, the energy sphere of $\nu_x$ is located at much smaller radius than $\nu_e$ and $\bar{\nu}_e$ due to the absence of charged current reactions, and therefore there are no reduction of $\nu_x$ number flux even in high energy neutrinos. For this reason, $\nu_x$ and $\bar{\nu}_x$ number fluxes become substantially larger than $\nu_e$ and $\bar{\nu}_e$. We note that neglecting the energy-exchanged scatterings and thermal processes for $\nu_x$ interactions result in overestimating neutrino energy flux in this study, but the qualitative trend that $\nu_x$ is more populated than $\nu_e$ in high energy region would hold even in cases with more realistic $\nu_x$ reactions.

\begin{figure*}
\begin{minipage}{1.0\textwidth}
    \includegraphics[width=\linewidth]{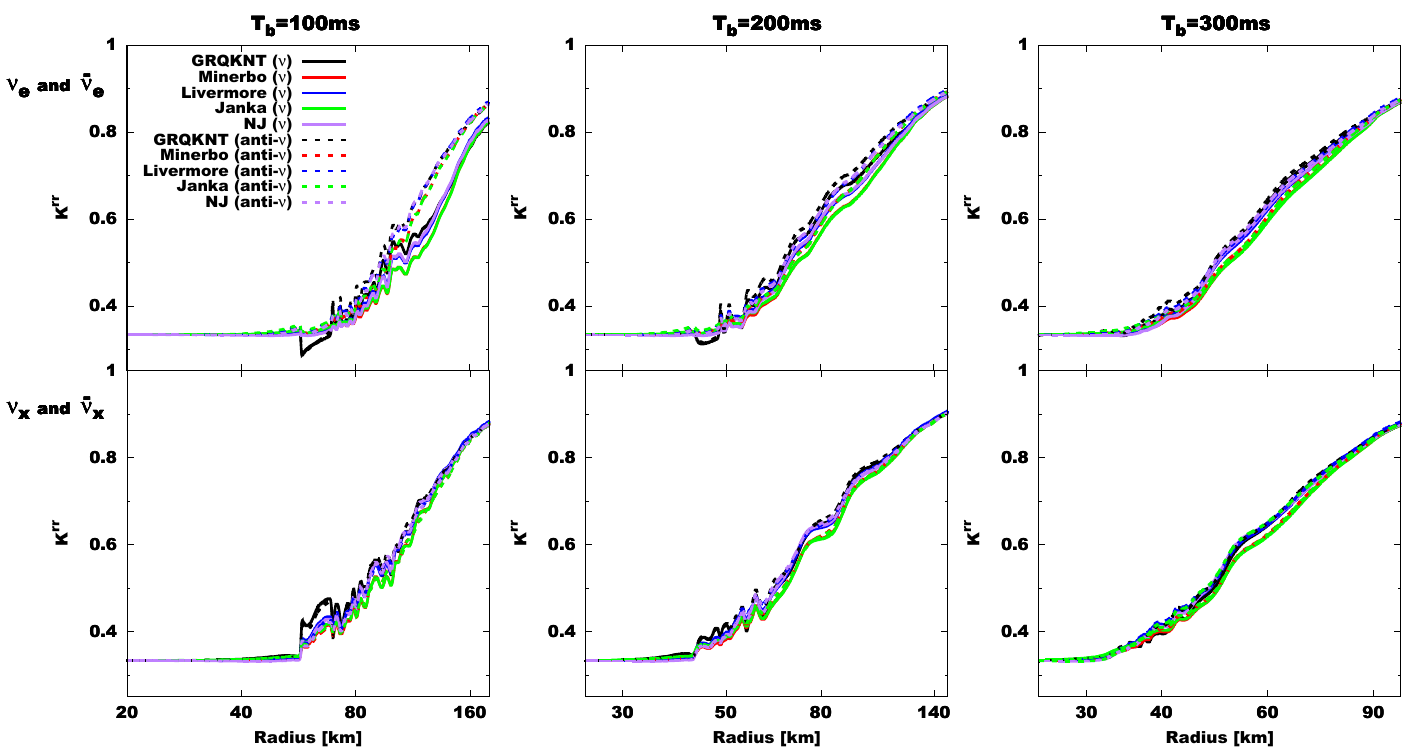}
\end{minipage}
    \caption{Radial profiles of $k^{rr}$ (r-r components of energy-integrated Eddington tensor) for diagonal components of density matrix. The top and bottom panels show the result of electron-type and heavy-leptonic neutrinos, respectively. We distinguish neutrinos and antineutrinos by line types, in which the former and latter are denoted by the solid and dashed lines, respectively. The colors distinguish various ways to compute Eddington tensors. Black lines show the Eddington tensors directly computed from full angular distributions (denoted as GRQKNT). For other colors, we compute Eddington tensor from zeroth and first angular moments (which are obtained from quantum kinetic simulations) with classical closure relations. In this study, we study four closure relations: Minerbo (red), Livermore (blue), Janka (green), and NJ (purple). From left to right, we display results for models of ${\rm T}_{\rm b}=100, 200$, and $300$ms, respectively.
}
    \label{graph_Closure_Krr}
\end{figure*}

\begin{figure*}
\begin{minipage}{1.0\textwidth}
    \includegraphics[width=\linewidth]{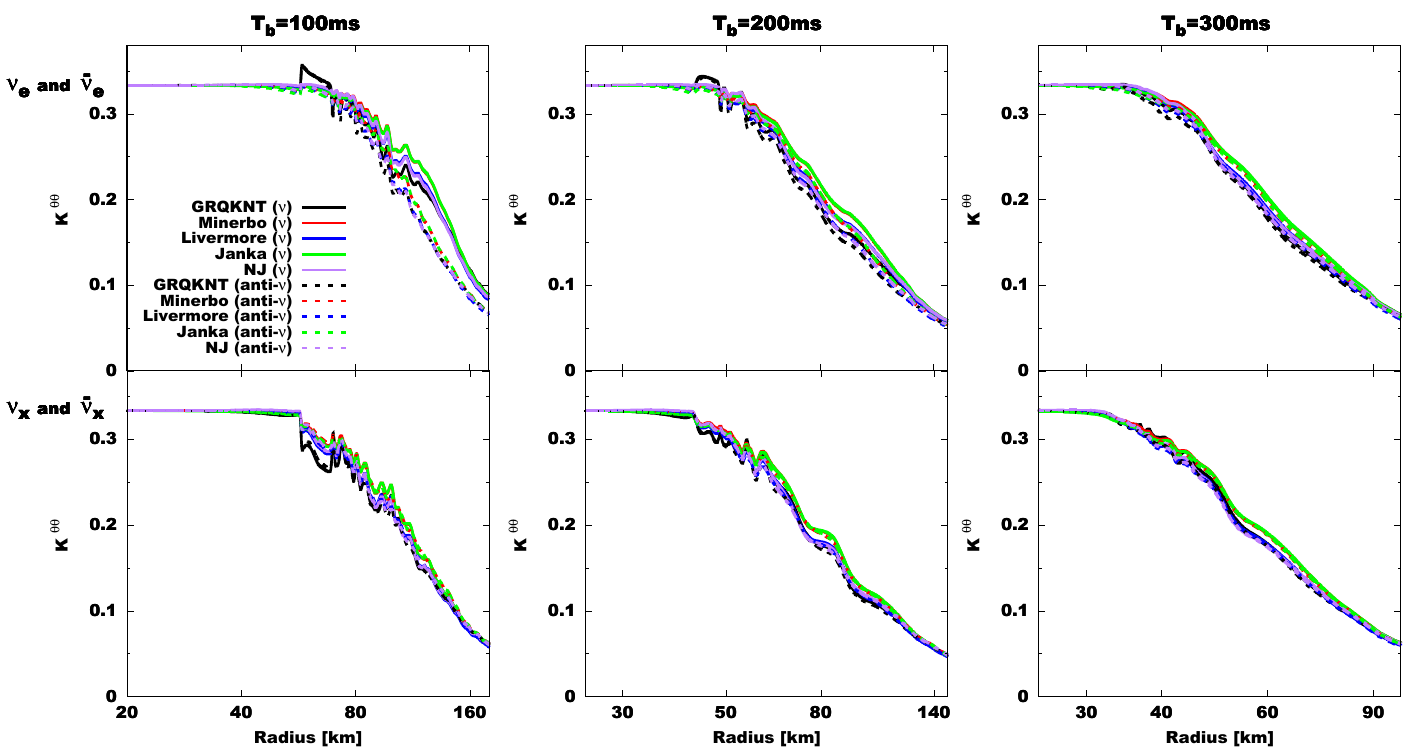}
\end{minipage}
    \caption{Same as Fig.~\ref{graph_Closure_Krr} but for $k^{\theta \theta} (= k^{\phi \phi})$.
}
    \label{graph_Closure_Kthth}
\end{figure*}

\subsection{Angular moments}\label{subsec:angmom}
Many neutrino transport schemes and associated neutrino radiation-hydrodynamic codes use some approximations to reduce the numerical cost in solving the transport equation. One of frequently used approaches is moment methods, in which the transport equation is reformulated in terms of angular moments. The modern approach is to solve only the zeroth and first angular moments, namely two-moment scheme. Although the angular integration in neutrino momentum space can substantially save the memory and computational time, truncating the moment hierarchy at a certain angular moments causes a non-trivial issue, so-called a closure problem. Determining accurate closure relations is not an easy task, because it hinges on not only local processes (e.g., neutrino-matter interactions) but also global geometries of the system (see, e.g., \cite{2020PhRvD.102h3017R,2021PhRvD.103l3025N}).

Recently, moment methods have also been proposed to solve QKE for neutrino transport \cite{2022PhRvD.105l3036M,2022arXiv220702214G}. They employed analytic closure relations which are essentially the same as those used in classical neutrino transport. It should be noted, however, that there are two caveats to use these classical closure approaches. First, determining Eddington tensors from the so-called flux factor is an ill-posed problem, since the zeroth angular moments for the off-diagonal components of density matrix are not positive definite. This implies that some special prescriptions need to be implemented; see, e.g., \cite{2022arXiv220702214G}. Second, there are no mathematically ill-posed problems to use these closure relations for diagonal components of density matrix, but we have not yet tested the validity of these classical closure relations. One concern regarding this issue is that neutrino angular distributions during flavor conversions are very complex, and small-scale structures are also developed (see, e.g., \cite{2020PhRvD.102j3017J}). Although we do not intend to account for the full detail of quantum closure relations, it would be very informative to present how well classical closure relations can approximate the Eddington tensors in quantum kinetic neutrino transport. For this purpose, we only focus on the diagonal components of density matrix in this analysis. Detailed studies towards developing quantum kinetic closure relations that can be applied to off-diagonal components are deferred to future work.

Figures~\ref{graph_Closure_Krr}~and~\ref{graph_Closure_Kthth} show the radial profiles of energy-integrated Eddington tensors obtained by various ways for the r-r and $\theta$-$\theta$ components, respectively. In this study, four classical closure relations are tested: Minerbo \cite{1978JQSRT..20..541M}, Livermore \cite{1984JQSRT..31..149L}, Janka \cite{1991ntts.book.....J}, and Nagakura and Johns (hereafter NJ) \cite{2021PhRvD.103l3025N}\footnote{The NJ closure is not analytic one but it is tabulated. The table is available from the link: https://hirokinagakura.github.io/scripts/data.html }. As shown in these figures, the variations of Eddington tensors obtained by different ways are small. Quantitatively speaking, Livermore and NJ closures offer reasonable values of Eddington tensors. We also find, however, that the Eddington tensors obtained by all classical closure relations have commonly large deviations from those computed from full angular distributions at $r \sim 60$ and $40$ km in models of ${\rm T}_{\rm b}=100$ and $200$ms, respectively. These spatial positions coincide with the transition layer where flavor conversions vigorously occur (see Fig.~\ref{graph_HeatCool_Aveene_Eneflux}). Below, we describe why classical closure relations can not approximate Eddington tensors well in the layer.

\begin{figure*}
\begin{minipage}{0.8\textwidth}
    \includegraphics[width=\linewidth]{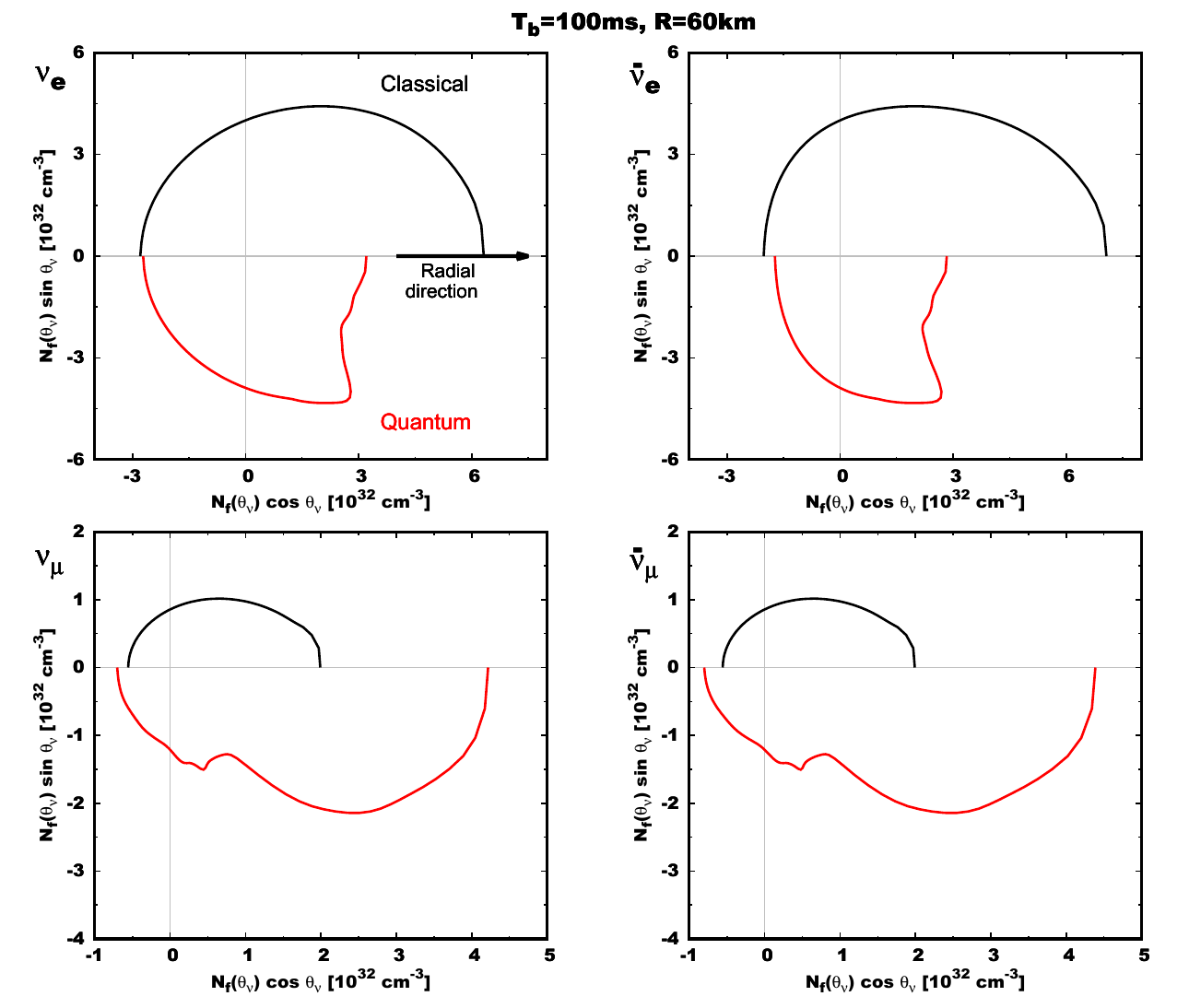}
\end{minipage}
    \caption{Energy- and azimuthally- integrated neutrino angular distributions at $60$km for ${\rm T}_{\rm b}=100$ms model. In each panel, we separate the results of classical neutrino transport (top) and quantum kinetic one (bottom) by a horizontal gray line. In the left top panel, we display the outgoing radial direction by a black arrow. Left and right panels distinguish results of neutrinos and antineutrinos, respectively, while top and bottom ones show results of electron- and $\mu$-type neutrinos, respectively.
}
    \label{graph_angNeutrino_t100ms_60km}
\end{figure*}

In the following, we focus on the energy- and azimuthally- integrated angular distributions of neutrinos, which is defined as,
\begin{equation}
N_{f} (\theta_{\nu}) \equiv \frac{1}{(2 \pi)^3} \int \int f E_{\nu}^2 d E_{\nu} d \phi_{\nu},
\label{eq:azim-ene-inte-distri}
\end{equation}
 at $r \sim 60$km for the model of ${\rm T}_{\rm b}=100$ms. As shown in Fig.~\ref{graph_angNeutrino_t100ms_60km}, strong flavor conversions occur for outgoing neutrinos. Both $\nu_e$ and $\bar{\nu}_e$ number densities and fluxes are larger than those of $\nu_x$ before flavor conversions occur. This results in a substantial depletion of $\nu_e$ and $\bar{\nu}_e$ by flavor conversions, and then reduce the fluxes, i.e., the first angular moments. For this reason, all classical closure relations fall into the illusion that the angular distribution approaches isotropic, and consequently the r-r component of Eddington tensor approaches $1/3$, despite the fact that the actual angular distribution is strongly anisotropic. As shown in Fig.~\ref{graph_angNeutrino_t100ms_60km}, the neutrinos at $\cos \theta_{\nu} \sim 0$ are largely populated compared to outgoing- and incoming neutrinos. These distributions are out of the range of applicability for all closure prescriptions tested in this study. In fact, the actual $k^{rr}$ becomes lower than $1/3$ (see black lines in left and middle top panels of Fig.~\ref{graph_Closure_Krr}), while $k^{rr}$ is always higher than $1/3$ in classical closure relations \footnote{This trend can also be qualitatively understood as follows. If the neutrinos are populated only at $\cos \theta_{\nu}=0$, both flux factor ($\kappa$) and $k^{rr}$ are exactly zero. However, the analytic closure relations gives $k^{rr}=1/3$, since the closure relation assumes that neutrino distributions are isotropic in the case with $\kappa=0$.}. For $\nu_x$, on the other hand, it is increased in all angles by flavor conversions, while the increase is more prominent in outgoing directions. This causes a sharp increase of $k^{rr}$ (see black lines in left and middle bottom panels of Fig.~\ref{graph_Closure_Krr}). Our result suggests that $\nu_x$ in $\cos \theta_{\nu} \sim 0$ in our multi-angle quantum kinetic simulations are less populated than those expected from analytic closure relations, and therefore multi-angle simulations provide higher $k^{rr}$ than those estimated from classical closure relations.

Contrary to $k^{r r}$, the abundant neutrinos at $\cos \theta_{\nu} \sim 0$ in $\nu_e$ and $\bar{\nu}_e$ results in the increase of $k^{\theta \theta}$. We find that $k^{\theta \theta}$ can be higher than $1/3$ in multi-angle quantum kinetic simulations. This is also beyond the applicability of classical analytic closures. On the other hand, $k^{\theta \theta}$ in multi-angle simulations for $\nu_x$ tends to be smaller than that expected from closure relations. This opposite trend from $\nu_e$ is in line with the result of $k^{r r}$.

Finally, we need to mention two important remarks. Although our result shows that the difference between closure-based Eddington tensors and those computed from multi-angle neutrino transport are small in most of the post-shock region, this does not guarantee that two moment schemes can always yield accurate results in quantum kinetic neutrino transport. The crucial condition in this test is that the zeroth and first angular moments are taken from the result of multi-angle simulations, i.e., the input is the same in this comparison. However, the dynamical simulations with two moment scheme should have, in general, some loss of accuracy in these low angular moments, which would become a potential source of error. Another remark in the present study is that we assess the closure relations based on time-averaged quantities. For more complete assessments, we need to extend our analysis to the time-dependent ones. In fact, the time-averaging neutrino distributions usually smear out small scale structures of flavor conversions. More work is definitely needed, and the improvement will be made in future under more consistent treatments of quantum closure relations.

\begin{figure*}
\begin{minipage}{1.0\textwidth}
    \includegraphics[width=\linewidth]{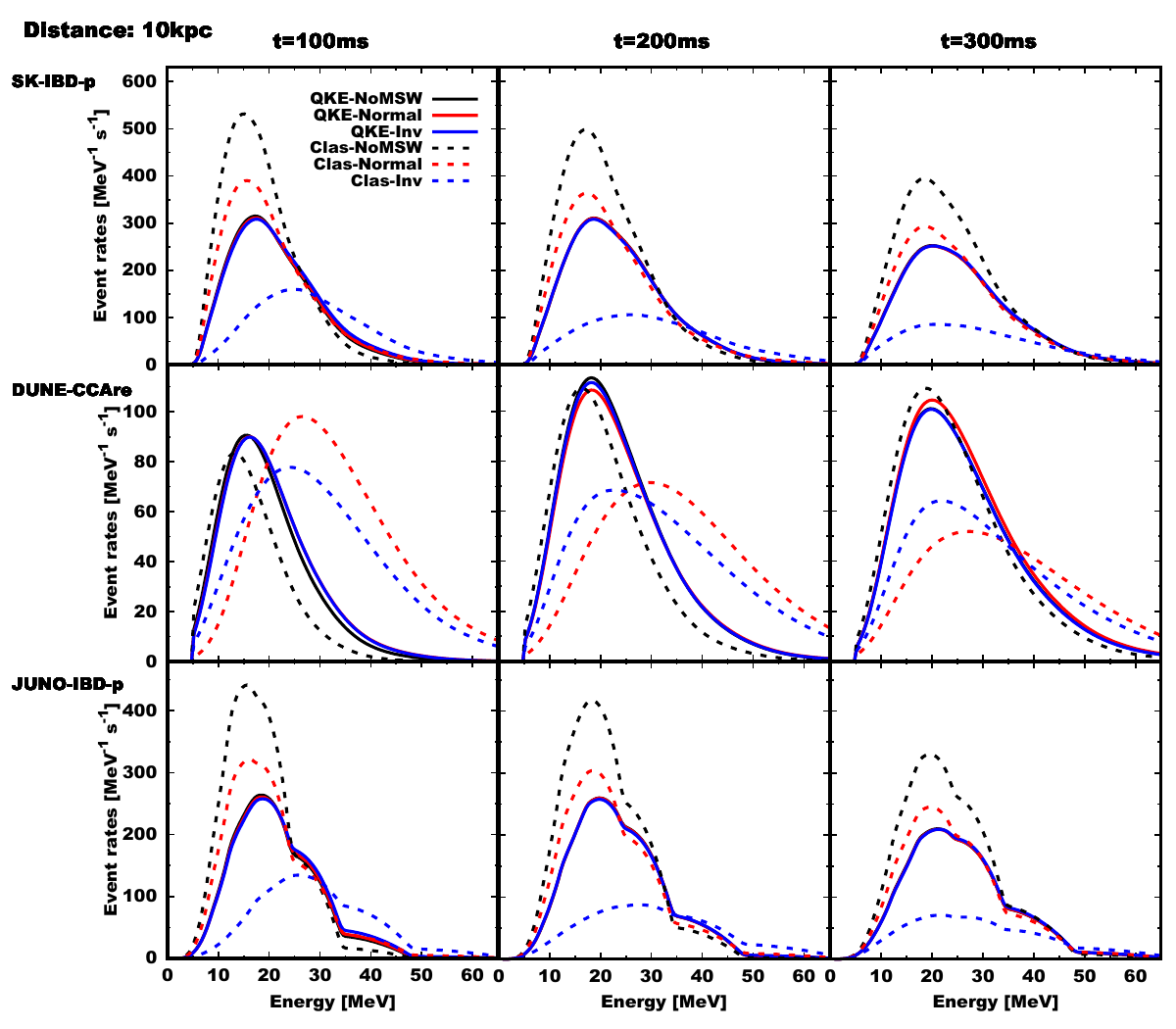}
\end{minipage}
    \caption{Energy spectra of neutrino event rate for the major reaction channel on each detector. From top to bottom, we show the results in SK, DUNE, and JUNO, respectively, while the results for models of ${\rm T}_{\rm b}=100, 200$, and $300$ms are displayed from left to right panels. Line type distinguishes the results between quantum (solid) and classical (dashed) simulations. We also take into account the adiabatic Mikheyev–Smirnov–Wolfenstein (MSW) neutrino oscillation effect, whose dependence is denoted by color: black (no MSW), red (normal mass hierarchy), and blue (inverted mass hierarchy). See the text for more details.
}
    \label{graph_DetectorSpect}
\end{figure*}

\subsection{Neutrino signal}\label{subsec:signal}
It is an intriguing question how the neutrino signal is altered by flavor conversions that occur inside neutrino spheres. We note that \citet{2023PhRvD.107j3034E} has also discussed the impact of flavor conversions on the neutrino signal. In their flavor mixing scheme, however, they imposed a condition of $\nu_x = \bar{\nu}_x$, which is not supported by our results; in fact, neutrinos and antineutrinos reach different asymptotic states in our simulations (see Figs.~\ref{graph_Radi_vs_angular_FFCtave_ELNmuLN}~and~\ref{graph_NumfluxSpect}). It is, hence, worthy to discuss impacts of flavor conversions on neutrino signal based on our results. As is the case in the study of angular moments (see Sec.~\ref{subsec:angmom}), however, we do not intend to analyze detailed features of neutrino signal.
This is because some simplifications adopted in the present study have an influence on the result. The detailed study, hence, should be postponed until more realistic CCSN models with high-fidelity neutrino-matter interactions are available. It should be mentioned, however, that some qualitative characteristics reported below would not be changed in more realistic situations.

In our detector simulations, we employ the same approach as used in \cite{2021MNRAS.500..696N,2021MNRAS.506.1462N}. The method is based on the SNOWGLOBES detector software\footnote{The software is available from the link: https://webhome.phy.duke.edu/~schol/snowglobes/}, that allows us to estimate the neutrino event rate and its energy spectrum for some representative neutrino detectors. In this study, we consider three detectors: SK, DUNE, and JUNO. We also note that SNOWGLOBES has a module to include effects of detector noise in the estimation, so these effects are taken into account in our result. On the other hand, we only focus on a major reaction channel in each detector, since the main purpose of this analysis is to discuss qualitative trends. Other subdominant channels would, however, play crucial roles in real observations, and we will discuss later how these channels can be used effectively in real observations.

The primary channel to detect CCSN neutrinos in SK and JUNO is to use inverse-beta-decay reactions on protons (IBD-p):
\begin{equation}
\bar{\nu}_e  + p \rightarrow e^{+} + n.
\label{eq:ibdp}
\end{equation}
In this study, the detector volumes in SK and JUNO are assumed to be $32.5$ and $20$ ktons, respectively. Since the detector configuration of Hyper-Kamiokande (HK) \cite{2018arXiv180504163H} is essentially the same as SK except for the volume, we can easily estimate the event rate in the case with HK by multiplying the SK event rate by a factor of 
$220/32.5 \sim 7$ \footnote{We note that the estimation is based on a HK design report at 2018 \cite{2018arXiv180504163H}, which may be changed in future.}.
 The dominant reaction channel in DUNE is, on the other hand, a charged-current reaction with Argon (CCAre):
\begin{equation}
\nu_e  + {^{40}{\rm Ar}} \rightarrow e^{-} + {^{40}{\rm K}^{*}}.
\label{eq:CCAre}
\end{equation}
The detector volume is assumed to be $40$ ktons. As shown in Eqs.~\ref{eq:ibdp}~and~\ref{eq:CCAre}, SK and JUNO are the most sensitive to $\bar{\nu}_e$, whereas DUNE has a capable of detecting $\nu_e$. We refer readers to \cite{2012ARNPS..62...81S} for more detailed information about properties of each detector.

In the estimation of neutrino event count, the dependence of Mikheyev–Smirnov–Wolfenstein (MSW) neutrino oscillation (which is mainly refractive effects by matter) is taken into account. This is distinguishable from neutrino flavor conversions in post-shock region, since this type of neutrino flavor conversion by MSW effects happens at the outer envelope of progenitors. One thing we do notice here is that, the three-flavor treatment of MSW effect is mandatory in the present study, and thus we follow the same approach in \cite{2021MNRAS.502...89N}. We adopt the neutrino mixing parameters in PMNS matrix from the result of y NuFIT 5.0 with SK atmospheric data \cite{2020arXiv200714792E}.
For the mixing angles, we set
$\theta_{12} = 33.44^{\circ}(33.45^{\circ})$, $\theta_{23} = 49.2^{\circ}(49.3^{\circ})$, $\theta_{13} = 8.57^{\circ}(8.60^{\circ})$ for normal (inverted) mass hierarchy.
See also Sec.4.1.2 in \cite{2021MNRAS.502...89N} for more details.

Figure~\ref{graph_DetectorSpect} summarizes our analysis, in which we show the energy spectra of event rate for a CCSN at a distance of $10$kpc. As shown the figure, noticeable features emerge in cases with quantum kinetic simulations. One of them is that the event rate is almost identical among different MSW cases. This is attributed to the fact that neutrinos have reached in flavor equipartitions before passing through the shock surface (but we stress that the equipartition states are different between neutrinos and antineutrinos; see also Sec.~\ref{subsec:basicpro}). This suggests that the neutrino event rate on each detector holds, no matter what neutrino oscillations happen during the neutrino propagation in the outer envelope of CCSN and also in the Earth \cite{2001NuPhB.616..307L}. It should be noted, however, that the event rate for high energy neutrinos ($\gtrsim 30$MeV) hinges on the neutrino oscillations models, since there are large differences in the energy spectrum of number flux between electron- and heavy-leptonic neutrinos (see Sec.~\ref{subsec:basicpro} for more details).


In real observations, flavor-dependent neutrino properties at the CCSN source will be retrieved from observed quantities. Let us finally discuss how we can determine whether observed neutrinos undergo large flavor conversions in the CCSN core in real observations. In conventional CCSN models, at least three independent reaction channels, having different sensitivities to different flavors of neutrinos, are necessary across multiple detectors to retrieve CCSN neutrinos \cite{2021MNRAS.500..319N}. This is attributed to the fact that there are three independent quantities in the neutrino fluxes: $\nu_e$, $\bar{\nu}_e$, and $\nu_x(=\bar{\nu}_x)$ at the CCSN source\footnote{Strictly speaking, this assumption is not valid since $\nu_x$ and $\bar{\nu}_x$ are not identical (even if on-shell muons do not appear in the CCSN core) due to high-order corrections of neutrino-matter interactions (as weak magnetism). However, the difference is expected to be minor except for high energy neutrinos (see also discussions of non-thermal neutrinos in \cite{2021MNRAS.502...89N}).}. Our result in the present study shows, however, that there are only two independent quantities in cases with large flavor conversions: $\nu$ and $\bar{\nu}$. This indicates that the observed quantities of DUNE and either SK or JUNO can provide the full information to retrieve the neutrino flux at the CCSN source.

We remark the consistency check for the above argument. This can be made by using other reaction channels which have the sensitivity to $\nu_x$, e.g., electron scatterings and neutral current reactions. Below, we give an explanation with an example. At first, we retrieve energy spectra of $\nu_e$ and $\bar{\nu}_e$ by CCAre channel of DUNE and IBD-p channel of SK or JUNO, respectively. By using the condition of flavor equipartitions, we can obtain the energy spectra for other heavy-leptonic neutrinos and anti-neutrinos\footnote{Let us stress again that this is an assumption, and strictly speaking there would be a disparity between $\nu_e$ and $\nu_x$ in high energy region (see Sec.~\ref{subsec:basicpro}). However, the majority of CCSN neutrinos have energy of $< 30$MeV, suggesting that the flavor equipartition is a reasonable approximation to discuss the overall trend. By using these energy spectra for all flavors of neutrinos, we can estimate the event count for other reaction channels having the sensitivity to heavy-leptonic neutrinos and antineutrinos, e.g., electron-scatterings and neutral-current reaction channels. We then compare the estimated event count to the actual observed one. If they are consistent with each other, the condition of $\nu_e = \nu_x$ and $\bar{\nu}_e = \bar{\nu}_x$ is reasonable, indicating that neutrinos and antineutrinos undergo strong flavor conversions in CCSN core.

There are two remarks about the consistency check. As described below, we need to estimate $\nu_e$ and $\bar{\nu}_e$ spectra accurately in this analysis, indicating that the joint analysis among DUNE, SK(HK), and JUNO, is mandatory. Second, it is necessary to estimate how large statistics are required in this analysis. As is well known, the observed data by subdominant detection channels would be noisy, suggesting that the threshold distance of CCSN for this consistency check should be considered. We leave this detailed investigation to future work.}

\section{Summary}\label{sec:sum}

Our understanding of many faces of neutrino quantum kinetics has deepened considerably in the last few years. However, there is still little known about how quantum kinetic features of neutrinos, represented as neutrino oscillations, affect the CCSN dynamics including explosion mechanism and its observable consequences. In this paper, we perform multi-energy, multi-angle, and three-flavor quantum kinetic neutrino transport simulations in spherical symmetry with a minimum but essential set of microphysics under fluid profiles taken from a CCSN model with full Boltzmann neutrino transport. Although some simplifications in neutrino-matter interactions and an artificial prescription as attenuating neutrino oscillation Hamiltonian are adopted, new insights into flavor conversions are obtained from the present study, which are summarized as follows.

\begin{enumerate}
\item Linear stability analysis suggests that collisional instability occurs in the optically thick region but it is overwhelmed by FFC in the regions where ELN angular crossings appear; see Sec.~\ref{sec:linana_ini} for more details.
\item ELN-XLN angular crossings disappear in the time-averaged neutrino profiles after the system reaches quasi-steady states in all quantum kinetic models, indicating that the disappearance of crossing is an intrinsic property of FFCs even with neutrino-matter interactions. However, the detailed asymptotic angular distributions of neutrinos are qualitatively different from those obtained in pure FFCs (no matter interactions), suggesting that neutrino-matter interactions play an important role in determining asymptotic states of neutrino radiation field.
\item Neutrino heating in the gain region is reduced by flavor conversions in all models, which is mainly due to the reduction of energy fluxes for $\nu_e$ and $\bar{\nu}_e$.
\item
On the contrary, the neutrino cooling in the optically thick region is enhanced, which is accounted for by increase of $\nu_x$ energy fluxes. These trends are consistent with our previous study in \cite{2023PhRvL.130u1401N}.
\item In quasi-steady state, flavor equipartitions are nearly achieved at large radii for all models. The equipartition state is, however, remarkably different between neutrino and antineutrinos.
\item We also find that heavy-leptonic neutrinos tend to be more abundant than electron-type ones in high energy region. This is indicative of occurrences of flavor conversions inside neutrino energy spheres. See Sec.~\ref{subsec:basicpro} for more details.
\item We assess the validity of classical closure relations for diagonal elements of density matrix of neutrinos, and find that the deviations from the correct Eddington tensors are small. On the other hand, some qualitative differences emerge in the region where flavor conversions vigorously occur. In this region, we show that Eddington tensors are out of the range of applicability for classical closure relations. See Sec.~\ref{subsec:angmom} for more details.
\item Neutrino signal becomes qualitatively different if strong flavor conversions occur inside neutrino spheres. On the other hand, flavor equipartition states, commonly observed in our quantum kinetic simulations, offer a potential strategy to identify the occurrence of strong flavor conversions around CCSN core in real observations. See Sec.~\ref{subsec:signal} for more details.
\end{enumerate}

In the future, we intend to explore more detailed features of flavor conversions in CCSNe by updating neutrino-matter interactions and with multi-dimensional CCSN models. These improvements will allow us to more detailed study for the impact of flavor conversions on CCSN dynamics. Aside from these practical issues, there still remain yet unexplored characteristics of flavor conversions, which should be addressed by fundamental studies of neutrino flavor conversions including analytic approaches combined with local simulations. This may further complicate the CCSN theory, but we will develop accurate prescriptions to accommodate these effects of flavor conversions in neutrino-radiation hydrodynamic simulations. The present study is a stepping stone for these future works.

\section{Acknowledgments}
We are grateful to Lucas Johns and Chinami Kato for useful comments and discussions. The numerical simulations are carried out by using "Fugaku" and the high-performance computing resources of "Flow" at Nagoya University ICTS through the HPCI System Research Project (Project ID: 220173, 220047, 220223, 230033, 230204, 230270), XC50 of CfCA at the National Astronomical Observatory of Japan (NAOJ), and Yukawa-21 at Yukawa Institute for Theoretical Physics of Kyoto University. For providing high performance computing resources, Computing Research Center, KEK, and JLDG on SINET of NII are acknowledged. This work is supported by High Energy Accelerator Research Organization (KEK). MZ is supported by a JSPS Grant-in-Aid for JSPS Fellows (No. 22KJ2906) from the Ministry of Education, Culture, Sports, Science, and Technology (MEXT) in Japan. HN is supported by Grant-inAid for Scientific Research (23K03468).
\bibliography{bibfile}

\end{document}